\documentclass[twocolumn,nopacs,prx]{revtex4-1}
\usepackage{amsmath}
\usepackage{amssymb}
\usepackage{graphicx}
\usepackage[bookmarks=false,linkcolor=blue,urlcolor=blue,colorlinks,citecolor=blue]{hyperref}
\usepackage[normalem]{ulem}
\usepackage{hyperref}
\usepackage{lipsum}
\usepackage{float}
\usepackage{bm}

\newcommand{\M}{\mathrm{M}}
\newcommand{\SM}{\mathrm{SM}}
\newcommand{\nm}{\mathrm{nm}}
\newcommand{\V}{\mathrm{V}}
\newcommand{\G}{\mathrm{G}}
\newcommand{\D}{\mathrm{D}}
\newcommand{\Al}{\mathrm{Al}}
\newcommand{\InAs}{\mathrm{InAs}}

\newcommand{\eV}{\mathrm{eV}}
\newcommand{\phd}{\phantom{\dag}}
\newcommand{\ph}{\phantom{.}}

\begin{document}

\title{Hybridization at superconductor-semiconductor interfaces}

\author{August E. G. Mikkelsen}
\author{Panagiotis Kotetes}
\author{Peter Krogstrup}
\author{Karsten Flensberg}
\affiliation{Center for Quantum Devices and Station Q Copenhagen, Niels Bohr Institute, University of Copenhagen, 2100 Copenhagen, Denmark}

\date{\today}
\begin{abstract}
{Hybrid superconductor-semiconductor devices are currently one of the most promising platforms for realizing Majorana zero modes. Their topological properties are controlled by the band alignment of the two materials, as well as the electrostatic environment, which are currently not well understood. Here, we pursue to fill in this gap and address the role of band bending and superconductor-semiconductor hybridization in such devices by analyzing a gated single Al-InAs interface using a self-consistent Schr\"odinger-Poisson approach. Our numerical analysis shows that the band bending leads to an interface quantum well, which localizes the charge in the system near the superconductor-semiconductor interface. We investigate the hybrid band structure and analyze its response to varying the gate voltage and thickness of the Al layer. This is done by studying the hybridization degrees of the individual subbands, which determine the induced pairing and effective $g$-factors. The numerical results are backed by approximate analytical expressions which further clarify key aspects of the band structure. We find that one can obtain states with strong superconductor-semiconductor hybridization at the Fermi energy, but this requires a fine balance of parameters, with the most important constraint being on the width of the Al layer. In fact, in the regime of interest, we find an almost periodic dependence of the hybridization degree on the Al width, with a period roughly equal to the thickness of an Al monolayer. This implies that disorder and shape irregularities, present in realistic devices, may play an important role for averaging out this sensitivity and, thus, may be necessary for stabilizing the topological phase.}

\end{abstract}

\maketitle

\section{Introduction}

Metal-semiconductor interfaces is a well-established field in the context of semiconducting electronics for implementing either Schottky diodes or Ohmic contacts, cf Refs.~\cite{SzeNgBook,Bardeen1947,Heine1965}. In the former case, the interface coupling forces the semiconductor to experience band ben\-ding leading to a depletion of charge near the semiconductor side of the interface, while the latter situation gives an accumulation of charges. The separation between the two beha\-viors is mainly determined by the sign of the difference of the metal work function ($W_{\M}$) and the electron affinity of the semiconductor ($\chi_{\SM}$), which we denote as $\Phi_\mathrm{bulk}=\chi_{\SM}-W_\M$. However, in practise other interfacial effect influence the band offset between the two mate\-rials. Here we work with an effective offset which we denote by $\Phi$. Recently, metal-semiconductor interfaces have attracted renewed attention in the context of engineered topological superconductivity where conventional $s$-wave Cooper pairing can be induced into the semiconductor. It has been theoretically predicted that the induced superconducting pairing combined with spin-orbit coupling (SOC) and an applied magnetic field can drive the system into the topological superconducting ($p$-wave)~\cite{BeenakkerReview,AliceaReview,LeijnseReview,AguadoReview,LutchynReview} state. This topologically non-trivial state supports charge-neutral zero-energy end states which obey non-Abelian exchange statistics. These so-called Ma\-jo\-ra\-na zero modes have interest for implementation of to\-po\-lo\-gi\-cal quantum computing~\cite{Nayak2008,Kitaev2001,Ivanov2001,Alicea2011,Aasen2016,Plugge2017,Karzig2017}.

The first experimental spectroscopic signatures of Majorana zero modes in superconductor-semiconductor hybrid devices were reported in Ref.~\citep{Mourik2012} and have since then been refined via the fabrication of ultra-clean epitaxial Al-InAs hybrids~\citep{Krogstrup2015,Chang2015}. These improved fabrication methods have led to promising reports of Majorana fingerprints in both epitaxial nanowire hybrids~\citep{Deng2016,Albrecht2016,Nichele2017,Zhang2018} and 2D epitaxial superconductor-semiconductor hybrids with lithographically defined 1D channels~\citep{Suominen2017,Nichele2017}.

\begin{figure*}[t!]
\includegraphics[width=\textwidth]{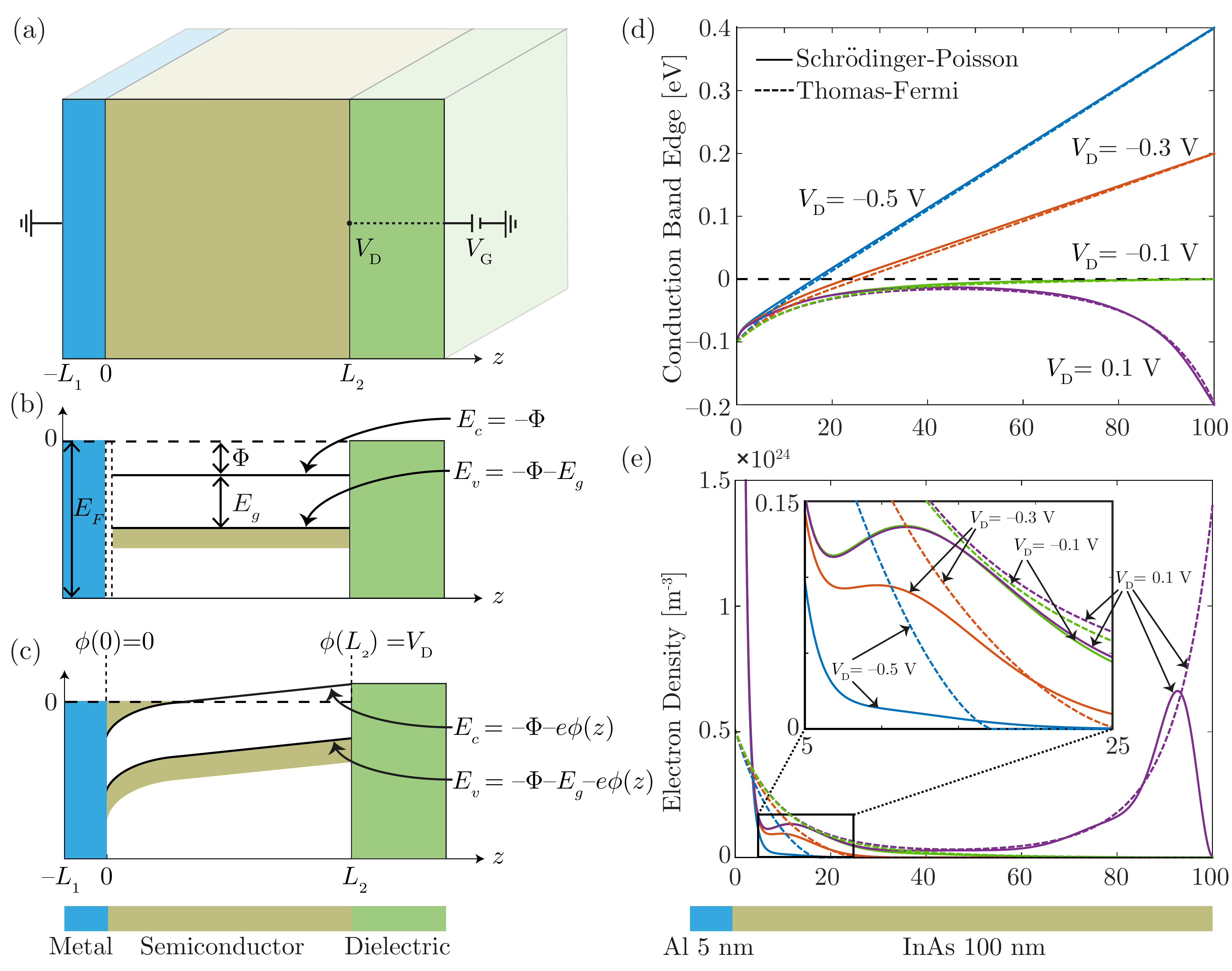}
\caption{(Color online) (a) Schematic of the hybrid device consisting of a layer of metal, semiconductor and dielectric with translational invariance in the plane. The rightmost edge of the dielectric is in contact with a gate electrode which keeps it at a voltage $V_{\G}$ with respect to the grounded metallic layer. This translates into a voltage difference between the grounded metallic layer and the lower edge of the semiconductor, which we denote $V_{\D}$. (b) Band diagram for the metal and semiconductor region before contact. (c) Band diagram of the metal and semiconductor region after contact. (d) Self-consistent band edge profiles in the semiconductor region for several values of $V_{\D}$. Solid lines indicate the results obtained via the self-consistent Schr\"odinger-Poisson method and dashed lines indicate the results obtained from the Thomas-Fermi approximation. (e) Self-consistent charge density profiles in the semiconductor region for several values of $V_{\D}$. Solid lines indicate the results obtained via the self-consistent Schr\"odinger-Poisson method and dashed lines indicate the results obtained from the Thomas-Fermi approximation.}
\label{fig:DeviceSketch}
\end{figure*}
\twocolumngrid

However, many microscopic details of the interfaces and in particular the degree of hybridization between the metal and semiconductor, which determines the induced superconducting pairing, the SOC strength and the effective $g$-factor, are not well-understood. Neither are the number of subbands occupied in the nanowires and the resulting semiconductor electron density. In fact, there seems to be a large spread in critical magnetic fields and gate voltages required to induce the topological state. A recent study showed a systematic dependence of the effective $g$-factor on gate voltage, when measured by the slope of the induced gap with applied field~\cite{Vaitiekenas2018}. Some the\-oretical progress in understanding the variation of the experimental results has already been made. For example in Ref.~\citep{Winkler2017} it was shown that orbital contributions might help to understand the large measured $g$-factor va\-lues, while the authors of {Ref.~\citep{Vuik2016,Dominguez2017,Escribano}} have assessed the role of the electrostatic environment in the nanowire devices. The effect of hybridization on the induced pairing has also been considered~\cite{VanHeck2016,Reeg2017}. However, the interplay between  hybridization and the electrostatically determined self-consistent potential has not been explored and a better understanding of this could potentially guide future experimental designs and theoretical modelling~\cite{Stanescu2017,Rainis2014,Prada2017} of Majorana devices.

In this paper we address the above-mentioned open issues by employing a self-consistent continuum model for the semiconductor-metal hybridization which incorporates band-bending effects due to electrostatics. Motivated by the existing epitaxial Majorana devices, we focus on an Al-InAs interface. Our analysis relies on applying the Thomas-Fermi and Schr\"odinger-Poisson me\-thods to calculate band edge and charge density profiles for the hybrid Al-InAs device shown in Fig.~\ref{fig:DeviceSketch}(a). The main conclusion of both approaches is that the band bending leads to a quantum well, which confines the electrons in the system in the vicinity of the superconductor-semiconductor interface. From the Schr\"odinger-Poisson approach we obtain the reconstructed band structure and investigate its response to varying the gate vol\-ta\-ge as well as the thickness of the Al layer. Our numerical findings are further understood via employing an analytical approach.

Based on our numerical investigation and approximate analy\-ti\-cal expressions, we discuss the hybridization and the resulting $g$-factor, SOC strength and induced superconducting gap for the interfacial bands crossing the Fermi energy. We find that the degree of hybridization for the bands with strong Al-InAs mixing is very sensitive to the thickness of the Al layer and native band offset $\Phi$, while only a weak dependence on the gate vol\-ta\-ge is found. In contrast, the hybrid bands with pre\-do\-mi\-nant InAs character are more susceptible to gating.

The above-mentioned pa\-ra\-me\-ters may be tuned such that a single band with strong superconductor-semiconductor hybridization crosses the Fermi level while leaving out bands with negligible coupling to the superconductor. However, the sensitivity to the Al width found here implies that a fine balance of parameter va\-lues may be required to obtain Majorana zero modes. In fact, this sensitivity appears even for Al thicknesses much larger than the ones rou\-ti\-nely employed in expe\-ri\-ments, e.g. $\sim10\nm$. Nevertheless, the dependence of the hybridization degree on the Al width, for the bands with mixed character, exhibits an alternating pattern ran\-ging from high to low values. The period of this pattern is approximately equal to the thickness of a single atomic Al layer. At first sight, this appears to contradict the claimed experimental observation of Majorana zero modes \cite{Mourik2012,Deng2016,Albrecht2016,Nichele2017,Zhang2018,Saulius}, in which, such a fine tuning is not yet accessible to such a degree. Nonetheless, additional effects to the ones considered here may account for this discre\-pan\-cy. First of all, even the highest fabrication quality devices exhibit shape imperfections stemming from residual strain in the semiconductor or Al-deposition irregularities. Evenmore, Majorana experiments are carried out in confined geometries, in which the breaking of translational invariance mimics the role of disorder. Therefore, we expect that the combination of the above disorder sources could ave\-ra\-ge out the periodically varying degree of hybridization.

\section{Self-consistent band bending}\label{sec:Section2}

\subsection{Setup and Electrostatics}

We consider the hybrid device depicted in Fig.~\ref{fig:DeviceSketch}(a), showing a layered structure consisting of metal, semiconductor and dielectric with translational invariance in the $xy$ plane. The device is characterized by two boundary conditions for the electrostatic potential $\phi$: (i) the metal layer is assumed to be a grounded conductor with $\phi = 0$, (ii) the rightmost edge of the dielectric is in contact to a back gate with $\phi=V_{\G}$. For our self-consistent modeling we focus only on the metal and semiconductor region of the device, and in this case (ii) is replaced by a scaled back gate voltage $\phi=V_{\D}$ at the semiconductor-dielectric interface. In fact, depending on the choice of the dielectric layer, $V_\G$ and $V_\D$ may differ by very little~\cite{GateScaling}.

We wish to determine the conduction band edge and charge density profile in our device. The idea behind our approach is shown in the band diagrams of Figs.~\ref{fig:DeviceSketch}(b) and (c). We assume that the Fermi level of the metal layer (dashed line) sets the chemical potential of the hybrid system and choose this as our reference energy. The distance between the Fermi level of the metal and its conduction band edge is determined by its Fermi energy, $E_{F}$, which we set to the bulk Al value, i.e. $E_{F}=11.7\,\rm{eV}$~\citep{Ashcroft}. Before contact (Fig.~\ref{fig:DeviceSketch}(b)), the conduction band edge of the semiconductor is assumed to be below the Fermi level of the metal corresponding to a \emph{positive} difference between the electron affinity of the semiconductor and work function of the metal, {that is $\Phi > 0$}. The results turn out to be very sensitive to this value and here we start by studying the case when the value of this expe\-rimen\-tal\-ly not yet fully resolved parameter is $\Phi=0.1\,\rm{eV}$.

When the metal and semiconductor layer are contacted (Fig.~\ref{fig:DeviceSketch}(c)), the band edges of the semiconductor will bend due to the presence of charges that are transferred from the metal into the semiconductor conduction band. We assume that only the conduction band electrons contribute to the band ben\-ding in the semiconductor, thus disregarding the presence of the valence band electrons. This approach is valid for the low-temperature range of interest, where $k_\mathrm{B}T \ll E_{g}\approx0.418\,\rm{eV}$~\citep{Winkler}. Furthermore, we must restrict ourselves to back gate voltages where the valence band edge of the semiconductor stays below the Fermi level of the metal; otherwise the band ben\-ding would lead to the formation of an unwanted hole pocket near the semiconductor dielectric interface. As indicated in Fig.~\ref{fig:DeviceSketch}(c), this condition is satisfied as long as $eV_{\D}>-\Phi-E_g=-0.518\,\rm{eV}$, which thus defines a lower bound for values of $V_{\D}$ that we can apply.

We determine the band edge and charge density profiles for our device using both a Thomas-Fermi approximation and a self-consistent Schr\"odinger-Poisson method. Both of these methods rely on determining $\phi(z)$ through Poisson's equation, which is solved only in the semiconductor region of our device
\begin{equation}\label{eq:Poisson}
\frac{d}{dz}\left[\varepsilon_{r}\frac{d\phi}{dz}\right]=-\frac{\rho(z)}{\varepsilon_{0}}.
\end{equation}
Here $\varepsilon_{r}$ denotes the dielectric constant of the semiconductor, which we set to $\varepsilon_{r}=15.15$ corresponding to InAs, while $\rho(z)$ denotes the charge density of conduction band electrons. The boundary conditions for Eq.~\eqref{eq:Poisson} are the previously described electrostatic boundary conditions, i.e. $\phi(0)=0$ and $\phi(L_{2})=V_{\D}$.

\begin{figure*}[t!]
\includegraphics[width=\textwidth]{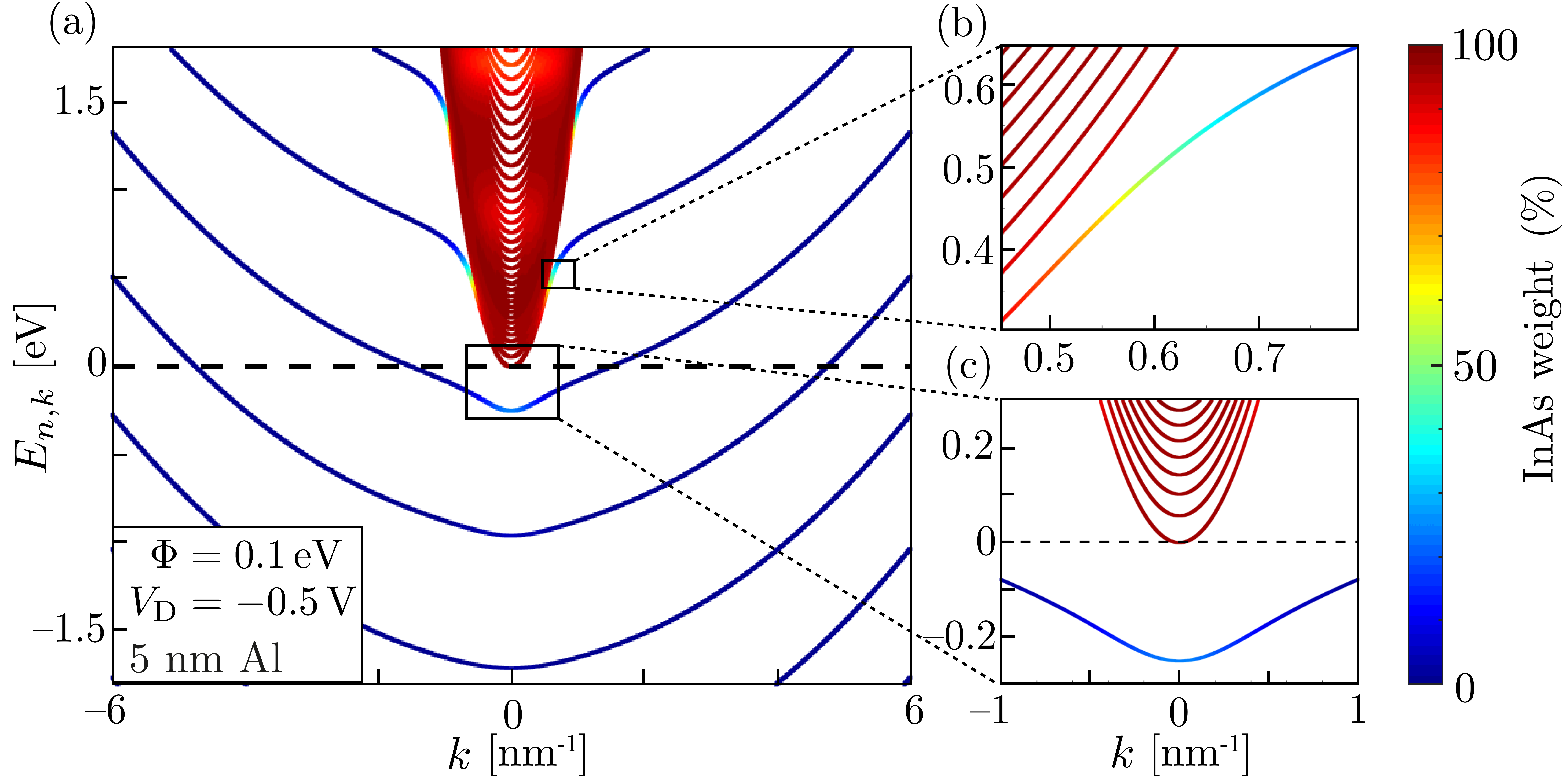}
\caption{(Color online) Hybrid band structure of Al-InAs obtained with parameters $L_{1}=5\,\rm{nm}$, $L_{2}=100\,\rm{nm}$, $\Phi=0.1\,\rm{eV}$ and $V_{\D}=-0.5\,\V$. (a) Large-scale zoom of the band structure showing that the bands are predominately Al-like with a narrow region of band segments, which have a large InAs weight. (b) Zoom-in showing that band segments with strong hybridization appear only for narrow ranges of $k$-values. (c) Zoom-in at the Fermi level revealing that this specific choice of parameters does not lead to states with strong hybridization at the Fermi energy.}
\label{fig:BandStructure1}
\end{figure*}

\subsection{The Thomas-Fermi approach}\label{sec:TF}

The Thomas-Fermi approximation relies on the assumption that the electronic charge density is given by the same expression as the standard result for a homogenous 3D electron gas
\begin{equation}\label{eq:ChargeDensityTF}
\rho(z) = -\frac{e}{3\pi^2}\left[\sqrt{2m_{\InAs}\epsilon_F(z)}/\hbar\right]^{3}.
\end{equation}
Here $\epsilon_{F}(z)$ denotes the local Fermi energy in the semiconductor, which is determined by the offset between the conduction band edge and Fermi level of the metal, i.e. $\epsilon_{F}(z)=\Phi+e\phi(z)$, while $m_{\InAs}$ denotes the effective mass of the semiconductor, which we set to $m_{\InAs}=0.023m_{e}$ corresponding to zincblende InAs~\citep{Winkler}.

The Thomas-Fermi approach further combines Eq.~\eqref{eq:ChargeDensityTF} with Poisson's equation ~\eqref{eq:Poisson}. This is most conveniently done via the introduction of the following rescaled quantites: (i) electrostatic potential $\varphi\equiv e\phi/\Phi$ and (ii) coordinate $\zeta\equiv z/\ell_{\rm TF}$. Here we defined the Thomas-Fermi length-scale $\ell^{-1}_{\rm TF}=\sqrt{e|\rho(\phi=0)|/(\varepsilon_{\rm InAs}\Phi)}$, where $\rho(\phi=0)$ denotes the charge density obtained in Eq.~\eqref{eq:ChargeDensityTF} for $\phi=0$. With the help of these rescaled quantities, Poisson's equation~\eqref{eq:Poisson} may be rewritten in the following dimensionless form
\begin{equation}\label{eq:PoissonDimless}
\frac{d^2\varphi(\zeta)}{d\zeta^2}=\big[1+\varphi(\zeta)\big]^{3/2}.
\end{equation}
The boundary conditions for $\varphi$ follows from the boundary conditions for $\phi$, i.e. $\varphi(0)=0$ and $\varphi(L_2/\ell_{\rm TF})=eV_{\D}/\Phi$. We solve Eq.~\eqref{eq:PoissonDimless} using standard numerical techniques for non-linear differential equations.

Figs.~\ref{fig:DeviceSketch}(d) and~(e) show the result of a Thomas-Fermi calculation (dashed lines) of the conduction band edge and charge density profile in an Al-InAs device for se\-ve\-ral values of $V_{\D}$. The calculation was done with thicknesses of $L_1=5\,\nm$ and $L_2=100\,\nm$ for the Al and InAs layers, respectively. We find that a triangular well forms near the Al-InAs interface due to band bending, which leads to charge accumulation close to the superconductor. Furthermore, for a positive value of $V_{\D}$, we find that electrons also accumulate near the dielectric away from the superconductor, thus leading to puddles with reduced superconductor-semiconductor hybridization. This situation is undesired since the hole puddle would give an ungapped region in the semiconductor and introduce quasiparticle poisoning to the Majorana device.

\subsection{The Schr\"odinger-Poisson approach}

The Schr\"odinger-Poisson approach is based on self-consistently solving Poisson's equation \eqref{eq:Poisson} together with the following Schr\"odinger equation
\begin{align}
-\frac{d}{dz} \left[ \frac{\hbar^2}{2m(z)}\frac{d\psi_{n,k}}{dz} \right] + \left[ E_{c}(z)+\frac{\hbar^2k^2}{2m(z)}\right]\psi_{n,k}  = E_{n,k}\psi_{n,k}.
\label{eq:Schrodinger}
\end{align}
The $xy$ plane translational invariance allows us to consider a fixed in-plane wavevector $\bm{k}=(k_{x},k_{y})$ of magnitude $k=|\bm{k}|$. The quantities $m(z)$ and $E_{c}(z)$ denote the effective mass and band edge of the hybrid system with the latter given by (see Fig.~\ref{fig:DeviceSketch}(c))
\begin{align}
E_{c}(z) &=
\left\{\begin{matrix}
-E_{F},  \quad& -L_{1}
\leq z
\leq 0, \\
-\Phi -e\phi(z), \quad&\quad 0 < z
\leq L_{2}.
\end{matrix}\right.
\label{eq:BandEdge}
\end{align}
The boundary conditions for the differential equation \eqref{eq:Schrodinger} are the hard-wall boundary conditions, i.e. $\psi_{n,k}(-L_{1})=\psi_{n,k}(L_{2})=0$. We solve it using a standard finite diffe\-rence approach, explained in Appendix~\ref{sec:Discretization}.

To obtain the self-consistent solution of Eqs.~\eqref{eq:Poisson} and \eqref{eq:Schrodinger} we need the electronic charge density which is found by integrating over the occupied eigenstates of Eq.~\eqref{eq:Schrodinger}, according to
\begin{equation}\label{eq:ChargeDensitySP}
\rho(z) = \frac{-e}{\pi} \int_0^{\infty} dk \  k  \sum_{{n}}\lvert \psi_{n,k}(z) \lvert^2 {\Theta(-E_{n,k})},
\end{equation}
with $\Theta$ denoting the Heaviside step function. We calculate $\rho(z)$ from the above by solving Eq.~\eqref{eq:Schrodinger} for many different values of $k$ and subsequently evaluating the integral numerically.

Figs.~\ref{fig:DeviceSketch}(d) and~(e) show the results of a Schr\"odinger-Poisson (solid lines) calculation of the conduction band edge and charge density profile with the same parameters as the previously described Thomas-Fermi calculation. The two approaches yield remarkably similar results for the band edge and quite similar results for the charge density profile. The strongest deviation for the latter appears at the metal-semiconductor interface, where the Thomas-Fermi result approaches the value predicted by Eq.~\eqref{eq:ChargeDensityTF}, while the value obtained using the Schr\"odinger method rises steeply due to the hybridization with the Al layer.

\begin{figure*}[t!]
\includegraphics[width=\textwidth]{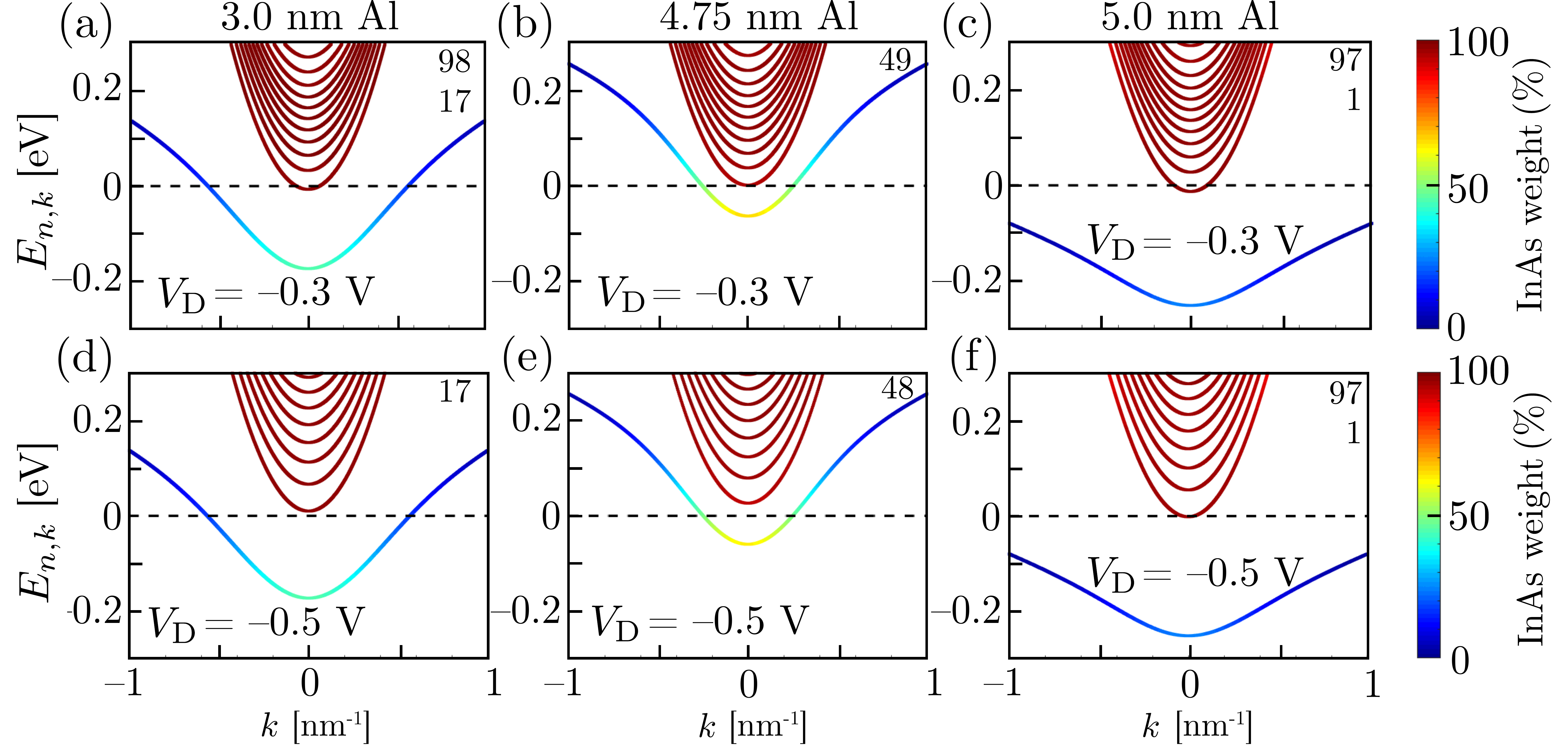}
\caption{(Color online) Band structure dependence on Al thickness and $V_{\D}$. Horizontal direction indicates changing Al thickness and vertical indicates changing $V_{\D}$. The semiconductor weights at the Fermi level of crossing bands are displayed in the top right corner of each sub\-fi\-gu\-re.}
\label{fig:BandStructure2}
\end{figure*}

\subsection{The hybrid band structure}\label{sec:BS}

Having determined the band bending profile, we proceed to investigate the hybrid band structure $E_{n,k}$ obtained from solving the Schr\"odinger equation \eqref{eq:Schrodinger}. Here we focus on results where $E_{c}(z)$ was obtained using the Schr\"odinger-Poisson approach, but in Appendix~\ref{sec:BandStructureComparison}, we also compare these to results based on simpler approximations including the Thomas-Fermi approach.

\begin{figure*}[t!]
\includegraphics[width=\textwidth]{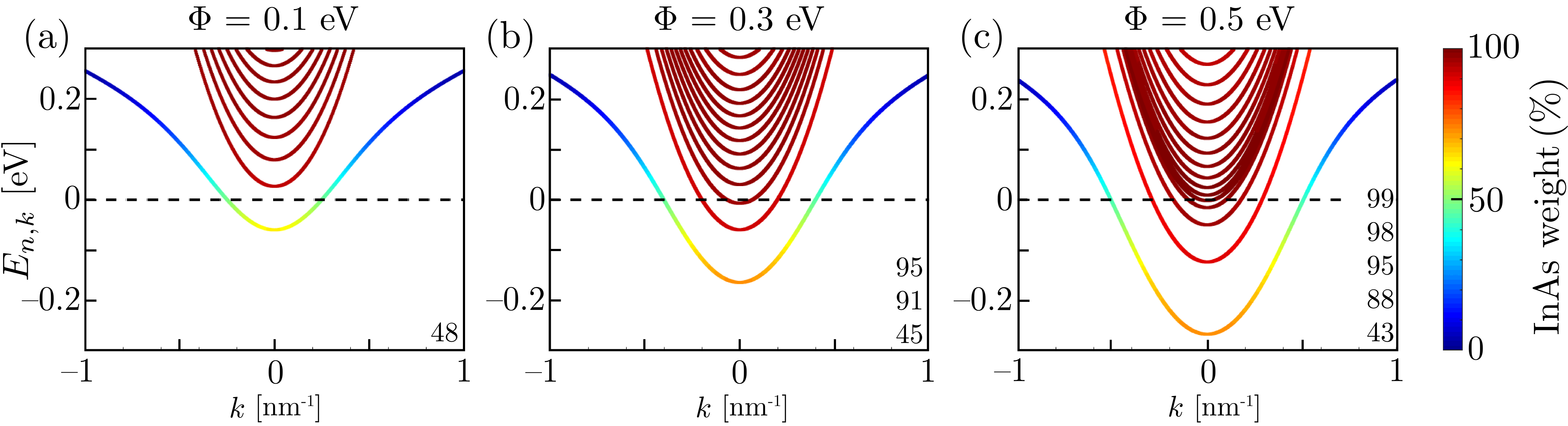}
\caption{(Color online) Band structure dependence on $\Phi$ obtained with parameters $L_{1}=4.75\,\rm{nm}$, $L_{2}=100\,\rm{nm}$ and $V_{\D}=-0.5\,\V$. Increasing the value of $\Phi$ leads to a deeper well at the superconductor-semiconductor interface, which gives rises to more InAs-like bands below the Fermi level. In (a,b,c) we show the band structure for $\Phi=0.1\,\rm{eV}$, $\Phi=0.3\,\rm{eV}$ and $\Phi=0.5\,\rm{eV}$, respectively.}
\label{fig:BandStructure3}
\end{figure*}

We first address the overall character of the bands based on the results displayed in Fig.~\ref{fig:BandStructure1} showing both $E_{n,k}$ as well as the weights in the InAs region of the corresponding wave functions $\psi_{n,k}$. {We have chosen to include negative values of $k$, such that the presented band structure corresponds to a cut through the 2D paraboloidal band structure of the system.} From the large-scale zoom provided by Fig.~\ref{fig:BandStructure1}(a) it is evident that the band structure is composed mainly of segments with negligible InAs weight corresponding to states localized in the Al region. It however also contains a dense region of band segments with high InAs weight corresponding to states which are predominantly localized in the InAs region. Strikingly, band segments with strong superconductor-semiconductor hybridization, i.e. with substantial weight of both InAs and Al, occur rarely as shown in Fig.~\ref{fig:BandStructure1}(b). We present in the next section an analytical approach for predicting when bands have substantial hybridization.

As discussed in Sec.~\ref{sec:SectionMF}, band segments with mixed weights are in fact crucial for the realization of robust Majorana zero modes. They correspond to states with both a strong SOC strength and sizeable supercon\-ducting gap, proportional to the Al character of the band, given by the weight $w_{\Al}$. Below we therefore investigate whether it is possible to obtain states of this character at the Fermi level. Furthermore, other bands that cross the Fermi level should not have large InAs weight ($w_{\InAs}$) at the Fermi energy since this would give rise to a soft gap.

To investigate the conditions for having states with mixed weights at the Fermi energy, we address the effects of varying the effective gate voltage $V_{\D}$ and Al layer thickness $L_1$, which are both parameters that can be tuned experimentally. Our results are summarized in Fig.~\ref{fig:BandStructure2}, where the top right corner of each sub\-fi\-gu\-re shows the weight at the Fermi level of the crossing bands. We focus on negative values of $V_{\D}$ for which all states are located close to the superconductor-semiconductor interface and restrict ourselves to gate voltages above $-0.518\,\V$ which, as previously discussed, defines a lower bound for values of $V_{\D}$.

Figs.~\ref{fig:BandStructure2}(b) and~(e) show that it is indeed possible to obtain a situation where a band with strong hybridization crosses the Fermi level, while InAs-like bands are kept above the Fermi level. Evidently this is obtained by tuning the Al layer thickness, which delicately determines the position and hybridization of the lowest Al-like band. In contrast, the hybridization of the bands responds more weakly to changes in $V_{\D}$. From Fig.~\ref{fig:BandStructure2} one can verify that lowering the gate vol\-ta\-ge has the expected effect of pushing up all the InAs-like bands, thereby depopulating states which would otherwise lead to a soft bulk gap. Nonetheless, gating appears only to affect the position of these bands, while their hybridization profile remains practically unchanged. In contrast, states with strong coupling to the superconductor essentially remain at the same energies. The reason for this is that the semiconductor component of the wavefunction of a strongly hybridized state is concentrated near the superconductor-semiconductor interface, where the superconductor screens the presence of the gate electrode.

To conclude this section, we investigate the effects of varying the parameter $\Phi$, which so far has been set to the value $\Phi=0.1\,\rm{eV}$. Our results are summarized in Fig.~\ref{fig:BandStructure3}, which displays the band structure at the Fermi level for different values of $\Phi$ using the parameters $L_{1}=4.75\,\rm{nm}$ and $V_{\D}=-0.5\,\V$ (chosen to achieve strong hybridization at the Fermi level). The results show that increa\-sing values of $\Phi$ lead to the emergence of more InAs-like bands below the Fermi level, thereby leaving se\-ve\-ral InAs-like states at the Fermi energy with negligible coupling to the superconductor. This is not surprising, since $\Phi$ effectively determines the depth of the quantum well at the superconductor-semiconductor interface, but it does appear contradictory to reported experimental measurements on hybrid Al-InAs structures which suggest a regime in which all states at the Fermi energy are strongly coupled to the superconductor. In our framework such a regime is achievable only with a small value of e.g. $\Phi = 0.1\,\rm{eV}$, which thus explains our motivation for originally choosing this value for $\Phi$. It should, however, be emphasized that this chosen value is somewhat smaller than found by recent angle resolved photoemission spectroscopy (ARPES) expe\-ri\-ments $\Phi \sim 0.23\,\rm{eV}$~\cite{ARPES}. The ARPES measurement were done for a bulk zincblende structure, but the relevant structure for the nanowire systems is wurtzite where the electron affinity is known to be $\sim$ 0.1 eV smaller. Therefore, using $\Phi\approx$ 0.1 eV for nanowire systems could be consistent with these expe\-ri\-ments. Moreover, it should be noted that these values significantly differ from the bulk value for the difference between the work function of Al and electron affinity of InAs~\citep{AlWorkfunction, InAsElectronAffinity} $\Phi_\mathrm{bulk}\sim 0.7\,\rm{eV}$.

\begin{figure}[t!]
\includegraphics[width=1\columnwidth]{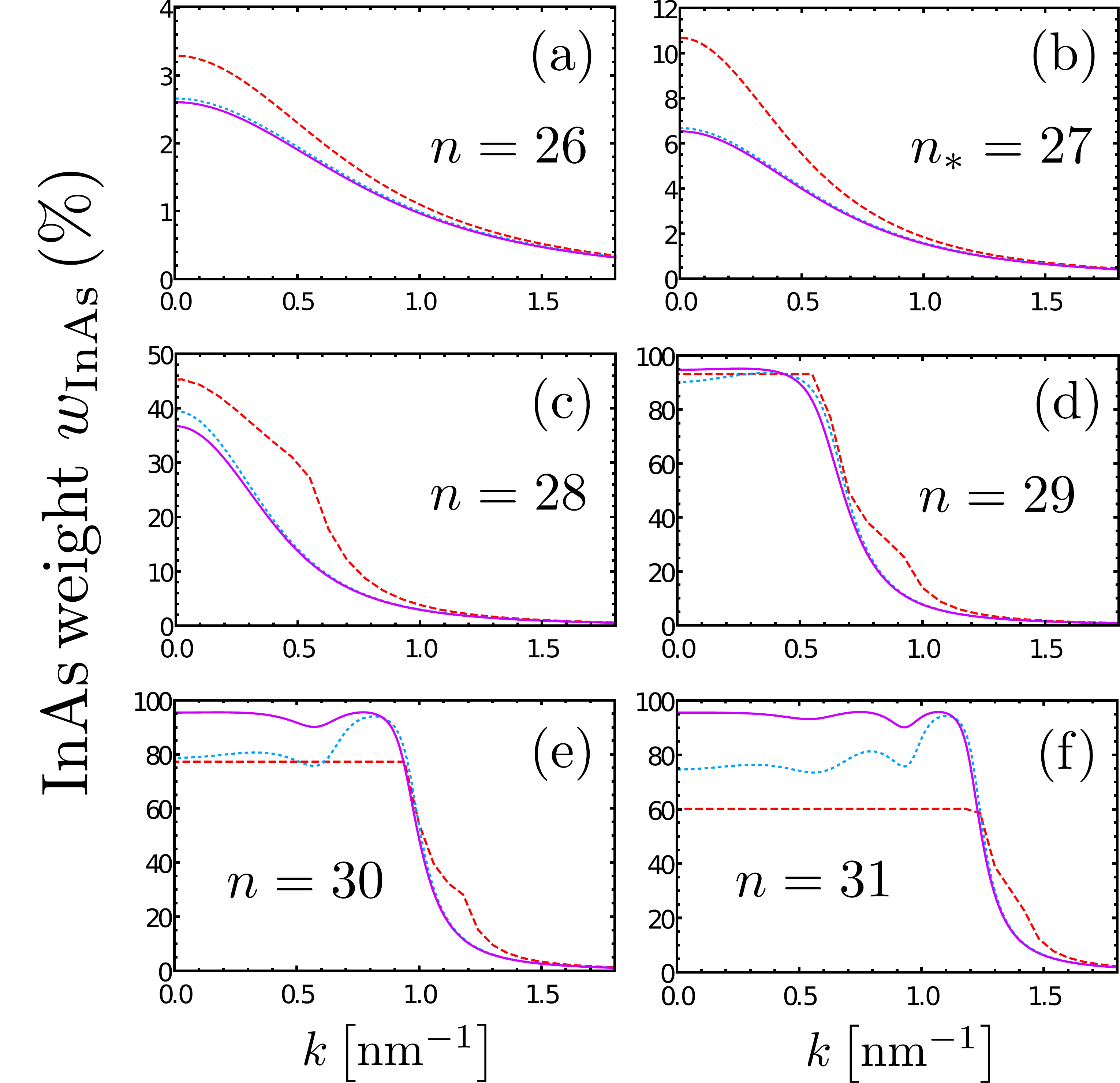}
\caption{(Color online) InAs weights for $L_1=5\,\nm$ and $\Phi=0.1\,\eV$. Solid purple line:~Weights obtained numerically using the self-consistent Schr\"odinger-Poisson method for $L_2=100\,\nm$. Dotted blue line:~Weights obtained numerically using the Schr\"odinger equation within the square-well model for $L_2'=16\,\nm$. Dashed red line:~Weights obtained within the square-well model using the appro\-xi\-ma\-te ana\-ly\-tical expressions of Eqs.~\eqref{eq:WAl},~\eqref{eq:Evawinas} and~\eqref{westimate} for $L_2'=16\,\nm$. The energetically-highest weakly mo\-di\-fied Al level is the one with $n_*=27$. We observe that the nu\-me\-ri\-cal methods are in good agreement up to $n=30$. For $n>30$ the nu\-me\-ri\-cal results obtained using the square-well show significant deviations from the ones calculated with the actual triangular potential. The analytical results follow the nu\-me\-ri\-cal ones and manage to capture the qualitative features of the weights. However, the approximate analytical approach is inadequate to describe bands with $n\geq31$.}
\label{fig:WeightsAl5nm}
\end{figure}

\section{Analytical approach to hybridization}\label{sec:Analytics}

\subsection{Effective-square versus triangular well model}

\begin{figure}[t!]
\includegraphics[width=1\columnwidth]{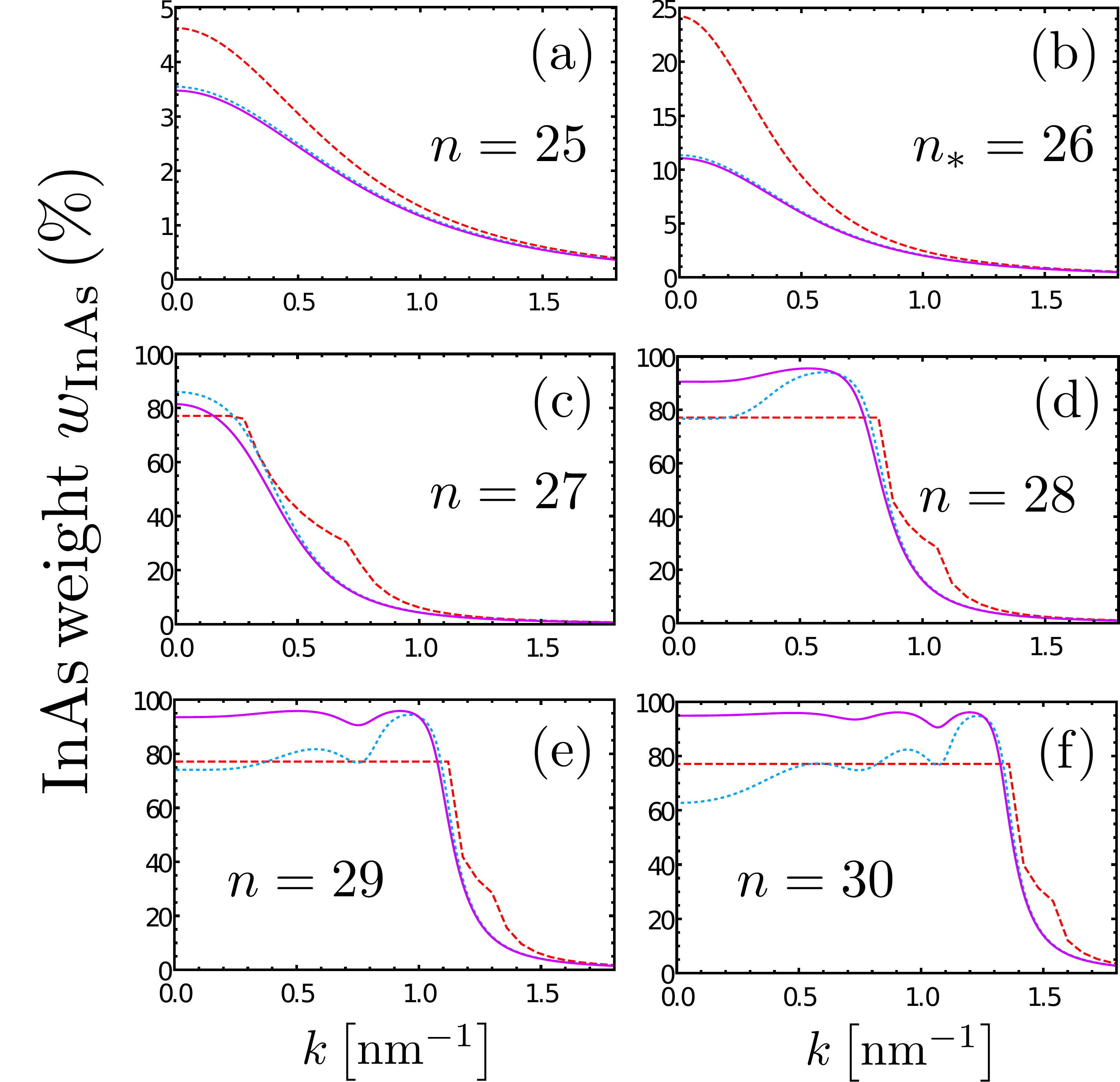}
\caption{(Color online) InAs weights for $L_1=4.75\,\nm$ and $\Phi=0.1\,\eV$. Solid purple line:~Weights obtained numerically using the self-consistent Schr\"odinger-Poisson method for $L_2=100\,\nm$. Dotted blue line:~Weights obtained numerically using the Schr\"odinger equation within the square-well model for $L_2'=16\,\nm$. Dashed red line:~Weights obtained within the square-well model using the appro\-xi\-ma\-te analytical expressions of Eqs.~\eqref{eq:WAl},~\eqref{eq:Evawinas} and~\eqref{westimate2} for $L_2'=16\,\nm$. The energetically-highest weakly mo\-di\-fied Al level is the one with $n_*=26$. The two numerical methods yield quite similar results up to $n=28$. For $n>28$ the numerical results obtained using the square-well show significant deviations from the ones calculated using the Schr\"odinger-Poisson method. The analytical results follow the nu\-me\-ri\-cal ones and manage to capture the qualitative features of the weights. The approximate analytical expression fail to describe weights corresponding to $n\geq29$.}
\label{fig:WeightsAl475nm}
\end{figure}

To shed light on the factors determining the degree of the superconductor-semiconductor hybridization, we proceed with studying a simpler and analytically tractable model. This model is obtained by replacing the triangular potential in the semiconductor with a rec\-tan\-gu\-lar well. This is applicable in the case of a large negative $V_{\D}$ and the situation corresponds to the one depicted in Fig.~\ref{fig:DeviceSketch}(b), with the only difference that the phy\-si\-cal width of InAs $L_2$ is replaced by an effective width $L_2'$, roughly given by the length for which the triangular potential crosses the Fermi level. In fact, by com\-pa\-ring the re\-con\-struc\-ted band structures of the two models, we find that they share the same qualitative features. To illustrate this connection, we compare in Fig.~\ref{fig:WeightsAl5nm} (\ref{fig:WeightsAl475nm}) the InAs weights obtained via the Schr\"odinger-Poisson method for parameters $\Phi=0.1\,\eV$, $L_2=100\,\nm$ and $L_1=5\,\nm$ ($L_1=4.75\,\nm$), with the ones calculated u\-sing the square potential for $L_2'=16\,\nm$. In the same plots we also include the weights calculated via employing the appro\-xi\-ma\-te analytic expressions to be discussed in the next paragraph. One observes that, given the way $L_2'$ is chosen, the weights obtained using these two models are in good agreement for energies near the Fermi level and begin to deviate for energies which lie above the Fermi level. This deviation mainly happens for small $k$ since the discrepancy is related to the sensitivity of the InAs-like bands to the electric field. The agreement allows us to extract approximate analytical expressions descri\-bing the hybridization characteristics using the square-well model.

\subsection{Square-well model and hybrid band structure}\label{sec:RecModel}

The wave functions in both regions are given by sinusoidal functions with wave numbers $k_\Al$ and $k_\InAs$. For a fixed $k$ they have the form
\begin{eqnarray}
\psi_{\Al}(z)&=&\frac{C_1}{\cal N}\sin[k_{\Al}(L_1+z)],\,\quad z\in[-L_1,0],\\
\psi_{\InAs}(z)&=&\frac{C_2}{\cal N}\sin[k_{\InAs}(L_2'-z)],\,\quad z\in(0,L_2'].
\end{eqnarray}
Due to the broken translational invariance along the $z$ direction, the wave numbers are determined by the ener\-gy
\begin{equation}
\frac{\hbar^2 k_\Al^2}{2m_\Al}+\frac{\hbar^2 k^2}{2m_\Al}-E_F=\frac{\hbar^2 k_\InAs^2}{2m_\InAs}+\frac{\hbar^2 k^2}{2m_\InAs}-\Phi=E.\quad\label{Edef}
\end{equation}
Here ${\cal N}$ and $C_{1,2}$ denote constants that will be found via the normalization and appropriate matching conditions at the interface $z=0$, respectively. The wave function matching yields the transcendental equation:
\begin{equation}\label{matchcondition}
\frac{m_\Al}{k_\Al}\tan(k_\Al L_1)=-\frac{m_\InAs}{k_\InAs} \tan(k_\InAs L_2').
\end{equation}
The above equation supports two types of solutions cha\-ra\-cte\-ri\-zed by: (i) $k_{\Al}\in\mathbb{R}$ and $k_{\InAs}\equiv i|k_{\InAs}|\in\mathbb{I}$ and (ii) $k_{\Al,\InAs}\in\mathbb{R}$. The first type describes dispersive solutions in Al which leak inside InAs within a width $\xi_\InAs=1/|k_{\InAs}|$. The second ty\-pe of solutions correspond to states which disperse in $z\in[-L_1,L_2']$.

\subsection{The band structure features}

Let us first discuss the bands which originate mainly from the pure Al bands. These become very weakly mo\-di\-fied by the InAs conduction band considered here, and belong to the first type of solutions mentioned earlier possessing an imaginary $k_{\Al}$. In fact, wave functions of this type will penetrate only a very small distance into the InAs region. For these deep bands, one has $k_\Al\approx n\pi/L_1$, corresponding to the Al layer being an infinite square well with energies
\begin{equation}\label{Enk}
E_{n,k}^{\Al}=\frac{\hbar^2k^2}{2m_\Al}+\frac{\hbar^2}{2m_\Al}\left(\frac{n\pi}{L_1}\right)^2-E_F.
\end{equation}
For the approximate energy dispersions of these bands see Appendix~\ref{sec:Analytix}.

There are $n_*$ such bands where $n_*$ is defined by
\begin{equation}
E_{n_*+\frac{1}{4},k=0}^{\Al}\lesssim-\Phi.
\end{equation}

\noindent For $\Phi=0.1~\eV$ and $L_1=5~\nm$ ($L_1=4.75\,\nm$) we find $n_*=27$ ($n_*=26$) which corresponds to $E_{n_*,k=0}^\Al\approx-0.76\,\eV$ ($E_{n_*,k=0}^\Al\approx-0.46\,\eV$). For these weakly mo\-di\-fied Al-like bands, the penetration depth into the semiconductor layer is approximately given by
\begin{equation}
\xi_\InAs^{-1}\approx\sqrt{k^2-2m_{\InAs}\left(E_{n,k}^{\Al}+\Phi\right)/\hbar^2}.\label{eq:xiAl}
\end{equation}
For the above parameters and $n_*=27$ ($n_*=26$), we find that $\xi_\InAs\approx1.6\,\nm$ ($\xi_\InAs\approx2.1\,\nm$) at $k=0$. At finite $k$ the penetration length becomes even smaller which becomes evi\-dent from the equation above. Finally, the InAs weight of such a $E_{n\leq n_*,k}$ band is approximately given by the expression
\begin{equation}
w_{\InAs}\approx\frac{(n\pi)^2}{(n\pi)^2+\left(\frac{m_{\Al}}{m_{\InAs}}\right)^2\left(\frac{L_1}{\xi_\InAs}\right)^3}.\label{eq:WAl}
\end{equation}
The numerical methods employed earlier for retrieving Figs.~\ref{fig:WeightsAl5nm} and~\ref{fig:WeightsAl475nm} confirm that the InAs weight increases with $n$, and for $n_*$ it attains values of the order of $5-10\%$. In the same figures we have also calculated the weight via the approximate Eq.~\eqref{eq:WAl}. We find that the InAs weight, while still reasonably low, is overestimated by this approximate formula. The same happens for $\xi_\InAs$.

\begin{figure*}[t!]
\includegraphics[scale=0.2]{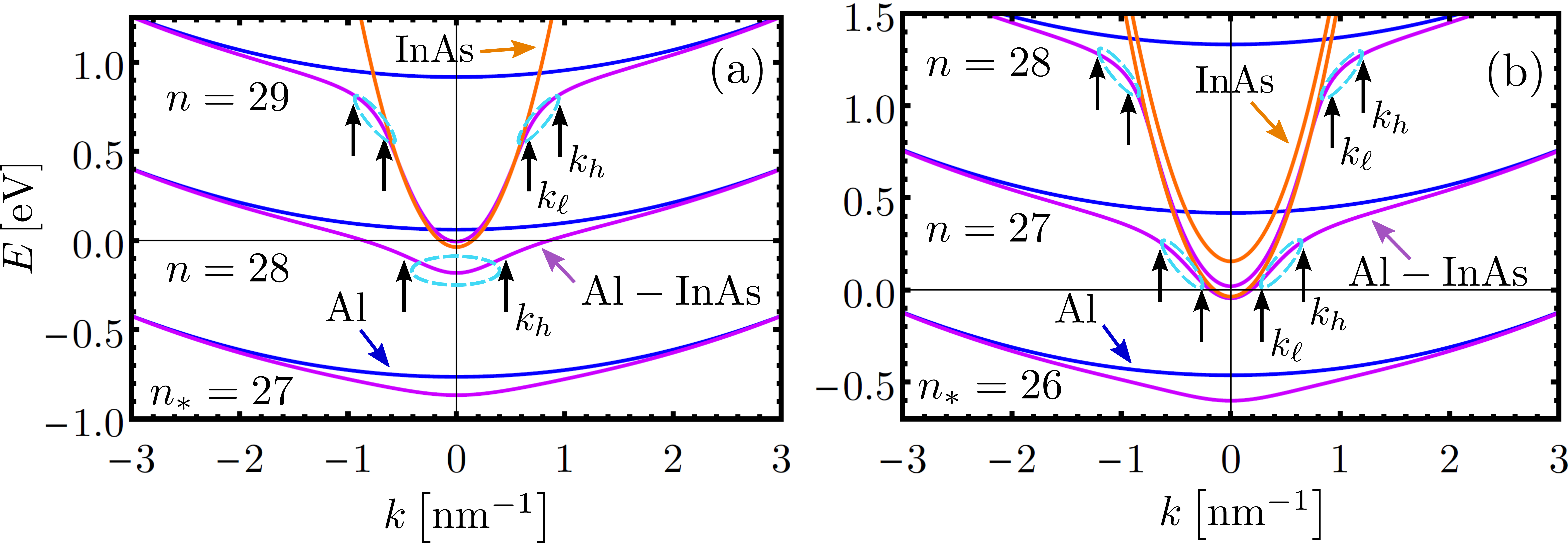}
\caption{(Color online) Figures (a) and (b) depict the two possible hybrid band structure scenarios for energies in the vicinity of the isolated InAs conduction band edge. The band structures shown are obtained via the self-consistent Schr\"odinger-Poisson method for $\Phi=0.1\,\eV$ and $L_2=100\,\nm$. In (a) ((b)) we have $L_1=5\,\nm$ ($L_1=4.75\,\nm$) and we have additionally shown the relevant pure Al and InAs bands for $L_2'=16\,\nm$. The dashed (cyan) ellipses illustrate band segments of strong superconductor-semiconductor hybridization which glue the pure Al and InAs bands together and generally appear for $k_\ell\leq k\leq k_h$, see Eq.~\eqref{eq:EvaCondition}. In (a) we encounter scenario \textbf{I} in which $k_\ell=0$ for $n=n_*+1$ and a band appears below the pure InAs bands due to the hybridization-induced downward bending of the $n_*+1$th pure Al band. This scenario is realized when the condition of Eq.~\eqref{EvaCondition} is satisfied. In contrast, in (b) this condition is not satisfied and a separated band does not appear below the pure InAs levels. However, gluing segments with $k_\ell>0$ appear so to connect the pure InAs and Al bands.}
\label{fig:Bandz}
\end{figure*}

In contrast to the low degree of hybridization achieved for the typical device parameters considered here when $n\leq n_*$, for $n>n_*$ it is possible to find bands that depending on the value of $k$ exhibit strong hybridization. We have studied the structure of the solutions of the transcendental equation~\eqref{matchcondition} and found that also the bands above $n_*$ have a one-to-one correspondence to the pure Al bands, in accordance with Ref.~\cite{Heine1965}. However, they significantly differ compared to the pure Al bands, since they become strongly modified in the presence of InAs, especially for small $k$. Therefore, a band with $n>n_*$ is generally divided into three $k$-space regions for which $k_{\InAs}\in\mathbb{R}$ or $k_{\InAs}\in\mathbb{I}$. See for instance Fig.~\ref{fig:Bandz}.

For large $k$ the new bands resemble the weakly mo\-di\-fied Al bands with $n\leq n_*$, since the pure InAs bands are concentrated in the small $k$ region. In most of the cases, the $n>n_*$ bands are thus Al-like as $k\rightarrow\infty$ and become InAs-like as $k\rightarrow0$. In between, new band structure segments appear upon hybridization,  which glue the pure Al and InAs bands together. These are characterized by $k_2\in\mathbb{I}$. Such segments are depicted in Fig.~\ref{fig:BandStructure1}(b,c) and illustrated with dashed (cyan) ellipses in Figs.~\ref{fig:Bandz}(a,b). In general, they appear for $k_{\ell}\leq k\leq k_h$, with $k_{\ell,h}$ given by the inequalities:
\begin{equation}
E_{n-\frac{1}{2},k}^{\Al}\leq\frac{\hbar^2k^2}{2m_{\InAs}}-\Phi\lesssim E_{n+\frac{1}{4},k}^{\Al}\quad{\rm for}\quad n>n_*.\quad\label{eq:EvaCondition}
\end{equation}

It is possible that $k_\ell=0$ and one obtains the situation of Figs.~\ref{fig:BandStructure1}(c),~\ref{fig:BandStructure2}(c) and~\ref{fig:Bandz}(a) of an Al-like band with additionally strong InAs character for small $k$. Essentially, the $n_*+1$th pure Al band is pushed downwards after contact with InAs, so that a new band now appears below the location of the pure InAs bands. See band $n=28$ in Fig.~\ref{fig:Bandz}(a). This occurs if the following condition is satisfied for the last Al-like band (meaning for $n=n_*$)
\begin{equation}
\label{EvaCondition}
E_{n_*+\frac{1}{2},k=0}^{\Al}\leq\frac{\hbar^2 k^2}{2m_{\InAs}}-\Phi.
\end{equation}

This type of gluing band segments, appearing only for $k\in[k_{\ell},k_h]$, possesses an InAs weight given approximately by (see Appendix~\ref{sec:Analytix})
\begin{equation}
w_\InAs\approx\frac{1}{1+L_1/\xi_\InAs}.\label{eq:Evawinas}
\end{equation}
For $n>n_*$ and $k\in[k_{\ell},k_h]$ the length $\xi_\InAs$ satisfies the approximate relation
\begin{align}
\label{winas}
1+L_1/\xi_\InAs\approx \sqrt{1+\pi^2\frac{m_\InAs}{m_\Al}\frac{\frac{\hbar^2 k^2}{2m_\InAs}-\Phi-E_{n-\frac{1}{2},k}^{\rm Al}}
{E_{1,k=0}^{\Al}+E_F}}.
\end{align}
Based on Figs.~\ref{fig:WeightsAl5nm} and~\ref{fig:WeightsAl475nm} one infers that it is precisely these band segments which have mixed Al-InAs character and exhibit strong hybridization. Thus, in order to ensure the robustness of the Majorana device one should maximize the difference $k_h-k_{\ell}$ and ensure that these segments cross the Fermi level. Strikingly, at the level of approximation considered here, the penetration depth, InAs weight and energy of these segments \textit{do not depend} on $L_2'$. This also implies a very weak dependence on the back-gate potential and $V_\D$, thus, explaining the findings of Fig.~\ref{fig:BandStructure2}.

The above discussion has already covered the bands with $n\leq n_*$ for all $k$, as well as the band segments for $n>n_*$ for $k>k_\ell$. We proceed with inve\-sti\-ga\-ting the properties of the band segments appearing for $k<k_\ell$ which primarily possess InAs character. At this point we distinguish two scenarios, corresponding to Figs.~\ref{fig:Bandz}(a,b), depending on whether a new band (scenario \textbf{I}) appears below the pure InAs bands or not (scenario \textbf{II}).

\textbf{Scenario I} When the condition of Eq.~\eqref{EvaCondition} is satisfied, a band appears below the pure InAs levels, as in Fig.~\ref{fig:Bandz}(a). In this case we find at higher energies hybridized InAs-like bands with $n>n_*+1$, $k<k_\ell$ and modified energies given by
\begin{equation}\label{EInAsmodified}
E_{n,k}\approx\frac{(n_*+1)^2E^{\InAs}_{n-n_*-1,k}+\lambda (n-n_*-1)^2E^{\Al}_{n_*+1,k}}{(n_*+1)^2+\lambda (n-n_*-1)^2},
\end{equation}
where we introduced the InAs energy levels
\begin{equation}\label{EInAs}
E_{s,k}^{\InAs}=\frac{\hbar^2k^2}{2m_\InAs}+\frac{\hbar^2}{2m_\InAs}\left(\frac{s\pi}{L_2'}\right)^2-\Phi
\end{equation}
and defined a hybridization coefficient
\begin{equation}\label{hybcoefdef}
\lambda=\left(\frac{m_{\Al}}{m_{\InAs}}\right)^2\left(\frac{L_1}{L_2'}\right)^3.
\end{equation}
The above approximation holds for bands satisfying $E_{n-n_*-1,k}^{\InAs}\lesssim E_{n_*+5/4,k}^{\Al}$, while for higher energies dif\-fe\-rent approximations apply. See Appendix~\ref{sec:Analytix}. The InAs weights for these bands are approximately given by
\begin{equation}\label{westimate}
w_\InAs\approx\frac{(n_*+1)^2}{(n_*+1)^2+\lambda (n-n_*-1)^2}.
\end{equation}
As it can be seen from Figs.~\ref{fig:WeightsAl5nm} and~\ref{fig:WeightsAl475nm}, the Al-content of these bands is practically negligible, at least for the parameter values considered here. {It is desirable that these band segments acquire a size\-able superconducting gap and get pushed to higher energies, and thus enhance the device protection against quasiparticle poi\-soning}. To achieve this goal one could use a metal with a smaller Fermi energy in order to reduce $n_*+1$.

\textbf{Scenario II} So far we have examined the situation in which the $n_*+1$th pure Al level does not get glued to a pure InAs level for $k=0$, but instead it is pushed downwards in energy yielding a band below the pure InAs ones. This is the case of Fig.~\ref{fig:Bandz}(a). {However, within the present model a slight modification of $L_1$ by $2.5\,{\rm \AA}$ can lead to a different situation in which the condition of Eq.~\eqref{EvaCondition} is \textit{not} satisfied. As a result one finds a different approximate expression for the InAs weight of the levels above $n_*$ and small $k$, i.e. $k<k_\ell$.} In this case, corresponding to Figs.~\ref{fig:BandStructure2}(b,e),~\ref{fig:BandStructure3}(a) and~\ref{fig:Bandz}(b), the pure Al levels become glued with the pure InAs ones via appropriate segments of mixed cha\-racter. While the InAs weight and $\xi_{\InAs}$ of these segments appearing for $k_\ell\leq k\leq k_h$ are given by the equations discussed earlier, the expressions describing the InAs-like parts living in $k<k_\ell$ become modified. We find that for $E_{n-n_*-1/2,k}^{\InAs}\lesssim E_{n_*+3/4,k}^{\Al}$ the energy of the segments defined for $k<k_\ell$ and $n>n_*$ read (see Appendix~\ref{sec:Analytix} for details)
\begin{equation}
E_{n,k}\approx\frac{L_2'E^{\InAs}_{n-n_*-\frac{1}{2},k}+L_1E^{\Al}_{n_*+\frac{1}{2},k}}{L_1+L_2'},\label{EInAsmodified2}
\end{equation}
and the InAs weight is given by the simple $(n,k)$-independent formula
\begin{equation}
w_\InAs=\frac{1}{1+L_1/L_2'}.\label{westimate2}
\end{equation}
We note that for $L_1/L_2'\approx 1/3$ the weight is $w_\InAs\approx75\%$. {From Fig.~\ref{fig:WeightsAl475nm} we find that this result agrees well with the corresponding Schr\"odinger-Poisson calculation for $n=27$. As $n$ increases one finds stronger deviations because for higher energies the differences between the square and triangular wells become more pronounced. In the next paragraph we show how to extend our approach and obtain an improved agreement with the Schr\"odinger-Poisson results.}

\subsection{Extended square-well model and fit to the Schr\"odinger-Poisson solution}

\begin{figure}[t!]
\includegraphics[width=.95\columnwidth]{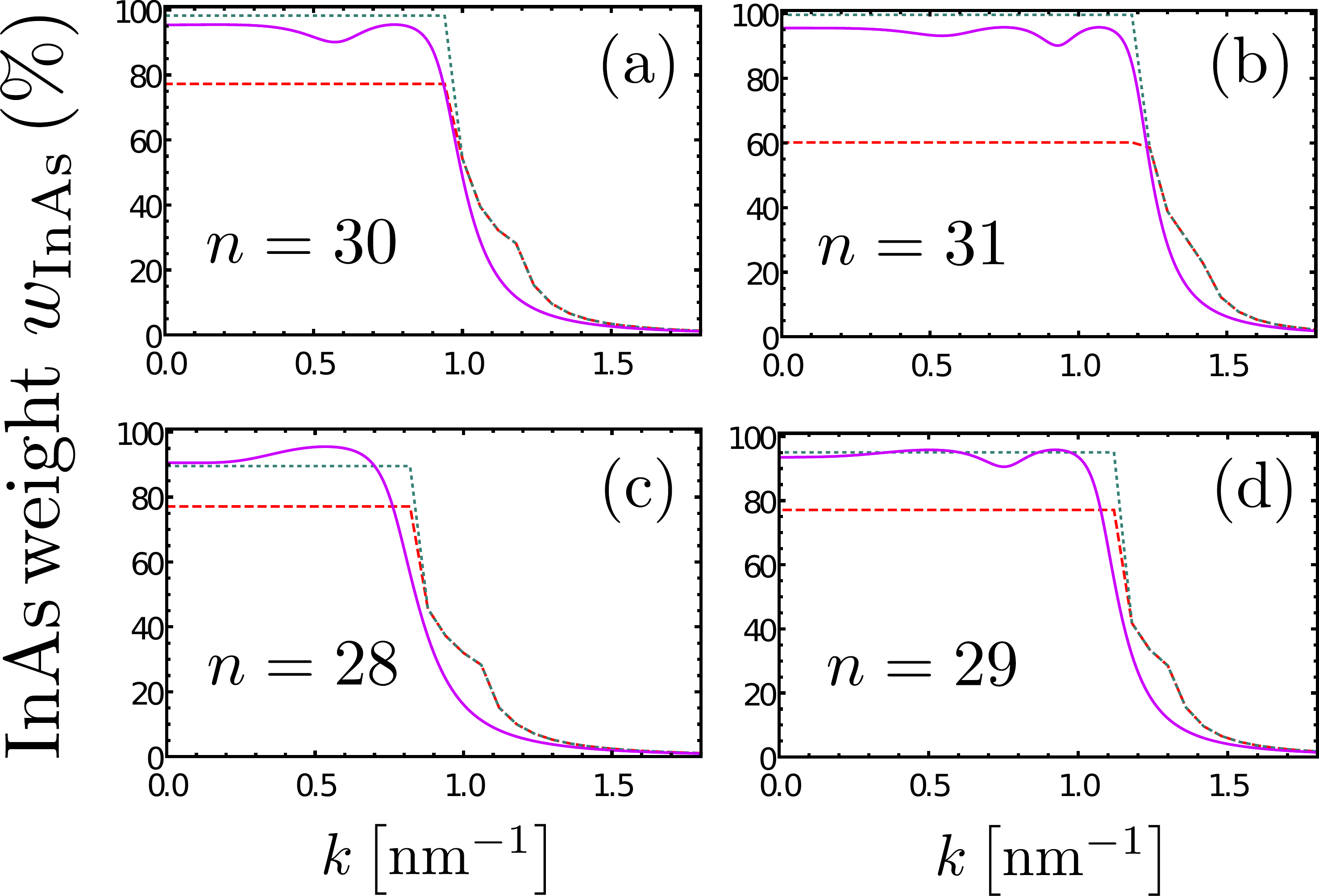}
\caption{(Color online) InAs weights shown in the upper (lower) panel for $L_1=5\,\nm$ ($L_1=4.75\,\nm$) and $\Phi=0.1\,\eV$. Solid purple line:~Weights obtained numerically using the self-consistent Schr\"odinger-Poisson method for $L_2=100\,\nm$. Dashed red line:~Weights obtained within the square-well model using $L_2'=16\,\nm$. Dotted green line:~Weights obtained using the extended square-well model which assumes an energy dependent $L_2'$ for bands above the Fermi level. Here we have used $L_2'=40.8\,\nm$ ($L_2'=91.6\,\nm$) for the left (right) panel. In both cases, the extended analytical model yields a significantly improved agreement with the Schr\"odinger-Poisson approach.}
\label{fig:ElectroWeights}
\end{figure}

The above conclusions can help us understand the obtained band structure when the electrostatic effects, introduced by a non-zero $\phi$, are taken into account. The strongly hybridized bands have a relatively short decay length inside InAs and weakly feel the electrostatic potential. On the other hand, the InAs-like bands are extended over a larger region and are prone to the gate-induced electric fields. The effect of the triangular well is to broaden the effective width $L_2'$ for energies above the Fermi level. In fact, for a band with index $n$ consisting of an InAs band with energy $E_{s,k}^{\InAs}\geq0$ ($s=n-n_*-1$ or $s=n-n_*$) one can define an effective energy dependent $L_{2,s}'$, given by
\begin{equation}
L_{2,s}'=\frac{E_{s,k=0}^{\InAs}}{e|{\cal E}_z|},
\end{equation}
since these bands appear for small $k$. Here ${\cal E}_z(z) = -d\phi/dz$ denotes the electric field in the system, which is non-zero only in the semiconductor's region. For $V_{\D}=-0.5~\V$ we find that $|{\cal E}_z|\approx 6.5~{\rm meV}\nm^{-1}$. By appropriately varying $L_2'$ depending on the band, we obtain a very good agreement with the Schr\"odinger-Poisson results as shown in Fig.~\ref{fig:ElectroWeights}.

\section{Effective parameters for Majorana devices}
\label{sec:SectionMF}

Having solved the electrostatics and studied the metal-semiconductor hybridization in detail, we now discuss its consequences for the realization of Majorana zero modes. The starting point for our analysis is the Bogoliubov-de Gennes Hamiltonian in the presence of the self-consistent potential determined above, see Eq.~\eqref{eq:BandEdge}, with added terms due to the Zeeman coupling, Rashba SOC and spin-singlet superconductivity. Using that the problem is translationally invariant in the $xy$ plane, we write the Hamiltonian in $\bm{k}$-space as
\begin{eqnarray}\label{eq:Hamiltonian}
&&{H}_{\bm{k}}(z)=\left[\hat{p}_z\frac{1}{2m(z)}\hat{p}_z
+\frac{\hbar^2k^2}{2m(z)}+E_{c}(z)\right]\tau_z\\
&&\quad+ \frac{g(z)\mu_B}{2}\bm{B}\cdot\bm{\sigma}+\alpha(z)\left(\hat{\bm{z}}\times\hbar\bm{k}\right)\cdot\bm{\sigma}\tau_z+ \Delta(z)\tau_x,\nonumber
\end{eqnarray}
where $\sigma_{x,y,z}$ ($\tau_{x,y,z}$) are Pauli matrices operating in spin (electron-hole space). Furthermore, $g(z)$ denotes the $g$-factor of the hybrid system, which is set to $g=+2$ in the Al region and $g=-14.9$~\citep{Winkler} in the InAs region. The magnetic field $\bm{B}$ is in-plane such that orbital effects play little role. The term $\alpha(z)$ is the Rashba SOC strength, which is given approximately by~\citep{Winkler}
\begin{equation}\label{SOCdef}
\hbar\alpha(z)=\frac{\hbar\bar{g}_{\rm InAs}\mu_B{\cal E}_z(z)}{2E_g}\equiv\frac{\bar{g}_{\rm InAs}}{2}\frac{\hbar^2}{2m_e}\frac{e{\cal E}_z(z)}{E_g},
\end{equation}
with
\begin{equation}
\bar{g}_{\rm InAs}=g_{\rm InAs}\frac{2m_e/m_{\rm InAs}+g_{\rm InAs}/2}{m_e/m_{\rm InAs}-g_{\rm InAs}/2}\simeq-23.3 .
\end{equation}
The final component entering Eq.~\eqref{eq:Hamiltonian} is the superconducting order parameter $\Delta(z)$ which is non-zero only in Al. For bulk Al, $\Delta_\Al\approx 340\,{\rm \mu eV}$.

If we assume large negative back gate voltages lea\-ding to a constant electric field in the semiconductor of the order of ${\cal E}_z(z)\approx 6.5\,{\rm meV}\nm^{-1}$, as estimated from the results in Fig.~\ref{fig:DeviceSketch}(d), the strength of the Rashba SOC in Eq.~\eqref{SOCdef} becomes $\hbar|\alpha_{\rm eff}|\approx 0.06w_\InAs\eV{\rm \AA}$. For $w_{\InAs}\approx0.5$ we find $\hbar|\alpha_{\rm eff}|\approx 0.03\,\eV{\rm \AA}$, $\Delta_{\rm eff}=w_\Al\Delta_\Al\approx170\,{\rm\mu eV}$ and $g_{\rm eff}=g_\Al w_\Al+g_\InAs w_\InAs\approx-6.5$.

\begin{figure}[t!]
\includegraphics[width=\columnwidth]{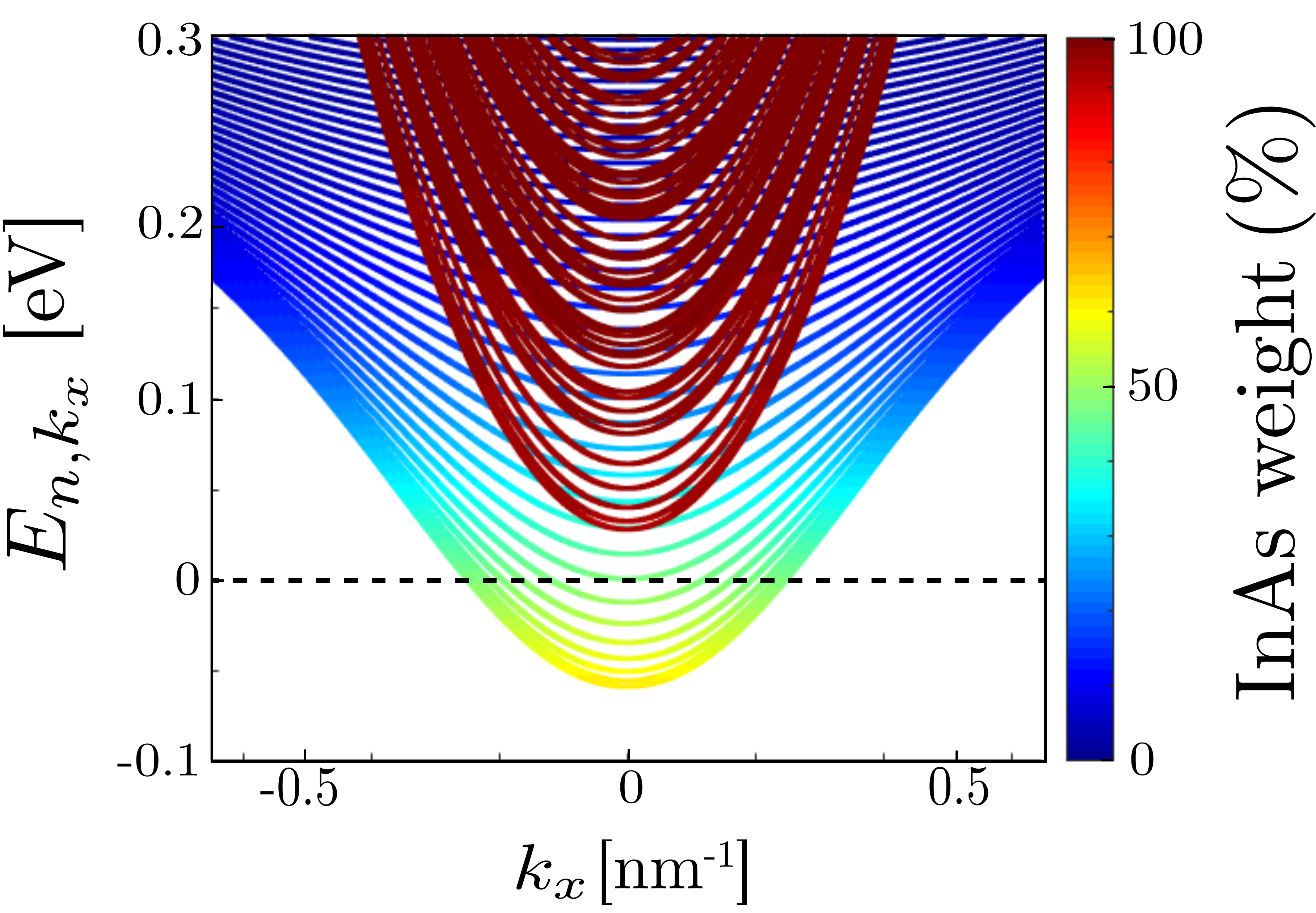}
\caption{(Color online) Hybrid bandstructure for an Al-InAs nanowire with a square $100\nm \times100\nm$ InAs cross-section. The band structure is obtained after imposing confinement along one of the planar directions (e.g. $y$). The parameter values are the same as in Fig.~\ref{fig:BandStructure2}(e). The confinement leads to a ``splitting'' of the original bands shown in Fig.~\ref{fig:BandStructure2}(e), thus, resulting into a finite number of channels which are shifted in energy but exhibit  similar hybridization profiles. We remark that, for presentation purposes, we have not included the confinement channels originating from purely metallic bands. Note also that we have neglected the possible electrostatic effects near the boundaries of the confined dimension.}
\label{fig:ConfinementBands}
\end{figure}

The effective superconducting gap, $g$-factor and che\-mi\-cal potential ($\mu_{\rm eff}$) enter into the condition deter\-mi\-ning the transition of an Al-InAs nanowire into the topological superconducting phase. When a single confinement channel of the nanowire hybrid band structure (see Fig.~\ref{fig:ConfinementBands}) crosses the Fermi level, the topological criterion reads $|g_{\rm eff}|\mu_B|\bm{B}|/2=\sqrt{\mu_{\rm eff}^2+\Delta_{\rm eff}^2}$. The dependence of these pa\-ra\-me\-ters on the InAs weight, for which we obtained ana\-ly\-tic expressions in the previous section, can be employed to indirectly infer the role of physical parameters such as the Al width, gate voltage and band offset. On the other hand, the effective mass ($m_{\rm eff}^{-1}=m_e^{-1} w_\Al+m_\InAs^{-1} w_\InAs$) and SOC strength determine the Majorana decay length $\xi_{\rm M}$, which further controls the resulting energy splitting and oscillations of overlapping Majorana zero modes in finite-sized nanowires~\cite{Cheng2009,Rainis2013,DasSarma2012,Mishmash2016}. The self-consistent Schr\"odinger-Poisson investigation of these effects in such a nanowire setup would require a full 3D simulation, which is a task beyond the scope of the present paper. Nevertheless, we can provide a rough estimate for $\xi_{\rm M}$. According to  Ref.~\onlinecite{Mishmash2016} we are in the case of weak SOC, which for the above parameter values implies $\xi_{\rm M}\propto \hbar/(m_{\rm eff}\alpha_{\rm eff})\approx0.5\ph{\rm \mu m}$. Finally, note that the inclusion of electrostatic effects can lead to a suppression or vanishing of the Majorana oscillations~\cite{Dominguez2017,Escribano}, through the zero-energy pinning of the state originating from the overlapping Majorana zero modes.

To this end, we note that in our self-consistent electrostatics analysis we neglected the SOC, Zeeman and superconducting contributions. We have verified that the inclusion of the SOC term in the self-consistent Schr\"odinger-Poisson problem does not sig\-ni\-fi\-cantly modify the obtained results. This also holds for both magnetic field and superconducting energy scales which constitute the smallest energy scales in the problem. Therefore, one could use the presented results for the band edge profiles and hybridization degrees for further modeling the physics of experimentally realized Majorana devices.

\section{Discussion and conclusions}

We have assessed the role of band bending and superconductor-semiconductor hybridization in Majorana devices by studying a planar, gated Al-InAs interface. Our results were based on a self-consistent Schr\"odinger-Poisson approach, which revealed that the band bending leads to an approximately triangular quantum well along with a charge accumulation layer at the Al-InAs interface. We also compared the Schr\"odinger-Poisson calculation with a Thomas-Fermi approach which ignores the hybridization and found remarkably similar results for the band bending. This can be useful for future calculations since one can use the computationally faster Thomas-Fermi approach to determine the self-consistent potential and then solve the Schr\"odinger equation in this potential.

The character of the superconductor-semiconductor hybridization was addressed by calculating the band structure of the hybrid system and investigating its response to varying the Al layer thickness, gate voltage and native band offset. Our main finding is that the system parameters may be tuned to a situation as shown in Fig.~\ref{fig:BandStructure2}(e) where a band with strong superconductor-semiconductor hybridization crosses the Fermi level, while higher levels of predominately InAs character stay above it. Such a situation is ideal for inducing superconductivity in the InAs region, which requires strong hybridization with the Al region, while simultaneously keeping out the InAs-like bands, which would give rise to a soft superconducting gap.

To back our numerical findings, we analyzed the superconductor-semiconductor hybridization using an analytical approach showing that the hybridization is only sizable when there is resonance between the uncoupled Al and InAs bands. This behavior might seem surprising given the absence of a barrier between the materials, and it appears to be a result of mismatch of the wavefunctions in the metal square well and triangular semiconductor well.

The conditions for ha\-ving the ideal situation shown in Fig.~\ref{fig:BandStructure2}(e) turns out to be extremely sensitive to the Al layer thickness, while a far weaker dependence on the gate voltage was found. As a matter of fact, in the regime of interest, the sensitivity to the Al width manifests itself through an alternating pattern of high and low values of the hybridization degree. Thus, the strong hybridization is not restricted to a single window of Al widths, but rather, it appears in a periodic fashion. An obvious question to address is whether the observed sensitivity persists for thicknesses much larger than $10\ph\nm$ which is a width ty\-pi\-cally employed in experiments. By investigating Al thicknesses such as 50nm and 60nm, we have found that a significantly better hybridization of the pure InAs-like bands can be achieved but that the observed sensitivity on the Al thickness remains. We expect that this sensitivity will persist, until the pure Al-level splitting for energies near the semiconductor's band edge is comparable to the splitting of the pure InAs-levels. For the parameters employed in the case of Fig.~\ref{fig:BandStructure2}(e) we find that $E_{n_*+1,k=0}^\Al-E_{n_*,k=0}^\Al\simeq\{83.6\ph {\rm meV},\,41.7\ph {\rm meV},\,27.8\ph {\rm meV}\}$ for the respective va\-lues of Al-width $L_1=\{50\ph \nm,\,100\ph \nm,\,150\ph \nm\}$. Thus, it appears that quite thick Al layers are required to soften the sensitivity in question. Note, that, in experiments small Al thicknesses are preferred in order to enhance the critical magnetic field at which the device becomes non-superconducting.

We have seen that the hybridization and the number of bands below the Fermi energy depend strongly on the thickness of the Al layer and therefore one expects that disorder such as variations by even a single monolayer of the deposited Al layer could have strong effects. Likewise, if a nanowire structure is formed out of the quasi-2D system studied here, a non-regular cross section could give qualitatively dif\-fe\-rent results from what we found for the 2D translationally invariant setup. Moreover, the strong dependence on Al thickness also raises the question whether a more microscopic description (for example a tight-binding model of the Aluminum) would give a different result. We speculate that the details will naturally be different but also that the sensitivity to thickness re\-mains because of the large mismatch of energy scales between the two ma\-te\-rials. Given the extreme sensitivity to the Al width, as discussed above, one may wonder how this is consistent with the experimental data which have been interpreted as signatures of induced topological superconductivity \cite{Mourik2012,Deng2016,Albrecht2016,Nichele2017,Zhang2018,Saulius}, because it would require atomically-flat Al along the whole length of the nanowire which seems unlikely to be case. However, the sensitivity may be softened by width variations on short length scales which could average out the hybridization degree. The length scale of roughening of the Al surface depends on growth conditions, semiconductor morphology and lattice matching. An example of a very highly ordered Al surface is when it is grown on lattice matched planar GaSb/InAs based materials where the interfacial domain matching with Al can be highly ordered. In this case the Al can follow the semiconductor surface morphology over several microns (verified by atomic force microscopy on structures with up to 300nm step size). But for most hybrid materials the roughness take place on much smaller length scales, down to the few nanometer scale. This roughness may be responsible for the proposed averaging. Therefore, both the effect of disorder and a more detailed band structure are natural questions for further research.

Another important parameter is how the metal Fermi level aligns with the semiconductor conduction band, see Fig.~\ref{fig:DeviceSketch}. Here we have used $\Phi$ in the range $0.1-0.3\ph \eV$, which is supported by recent experiments~\cite{ARPES}, but not by known bulk values. Therefore, it could be that there is some surface chemistry that still needs to be resolved before a more complete understanding of these structures can be reached.

In conclusion, devices based on Al-InAs or similar material combinations are indeed promising candidates for Majorana physics and several experiments have already shown signatures of Majorana zero modes. However, based on the analysis here it seems to require a fine ba\-lan\-ce between several parameters, such as the metal thickness and band alignments. In our simulations the effect on gate voltage is very limited when it comes to the degree of hybridization, while it predominantly affects the position of the InAs-like bands. Disorder effects might help in relaxing these conditions. Studying these effects would require a 2D simulation, which we intend to pursue in future works. Experimentally, there seems to be a stronger dependence on gate voltage which could be due to the gate coupling inhomogeneously to the structure. A better understanding of this would require a full 3D self-consistent simulation.

At the time of submission two other works addres\-sing Schr\"odinger-Poisson calculations for superconductor-semiconductor hybrid structures appeared~\cite{Antipov2018,Woods2018}. We find that our results on the hybridization are in a good agreement with those obtained in these works using numerical methods. The primary interest of Refs.~\onlinecite{Antipov2018,Woods2018} is to explore aspects of the topological phase dia\-gram of nanowires. Specifically, the authors of Ref.~\onlinecite{Antipov2018} discuss the influence of gating on the effective $g$-factor, while Ref.~\onlinecite{Woods2018} focuses on efficient nu\-me\-ri\-cal me\-thods for solving the Schr\"odinger-Poisson problem, as well as, on the electrostatic effects on the Majorana oscillations. In the present paper, we have put emphasis on understanding in detail the superconductor-semiconductor hybridization and investigated the sensitivity of the degree of hybridization upon varying key parameters (\textit{e.g.}, Al thickness). We have also derived approximate analytic expressions enabling experimental predictions.\\

\section*{Acknowledgements}

We would like to thank A.~Akhmerov, A.~Antipov, K.~Bj\"ornson, M.~Hell, R.~Lutchyn, S.~Schuwalow, S.~Vaitiek\'enas, A.~Vuik, G.~Winkler, and M.~Wimmer for useful discussions. This work was supported by the Da\-nish National Research Foundation, by the Microsoft Station Q Program and by the European Research Commission, project HEMs-DAM no.716655,
starting grant under Horizon 2020..

\appendix

\section{Numerical methods}\label{sec:Discretization}

\begin{figure*}[t!]
\includegraphics[width=\textwidth]{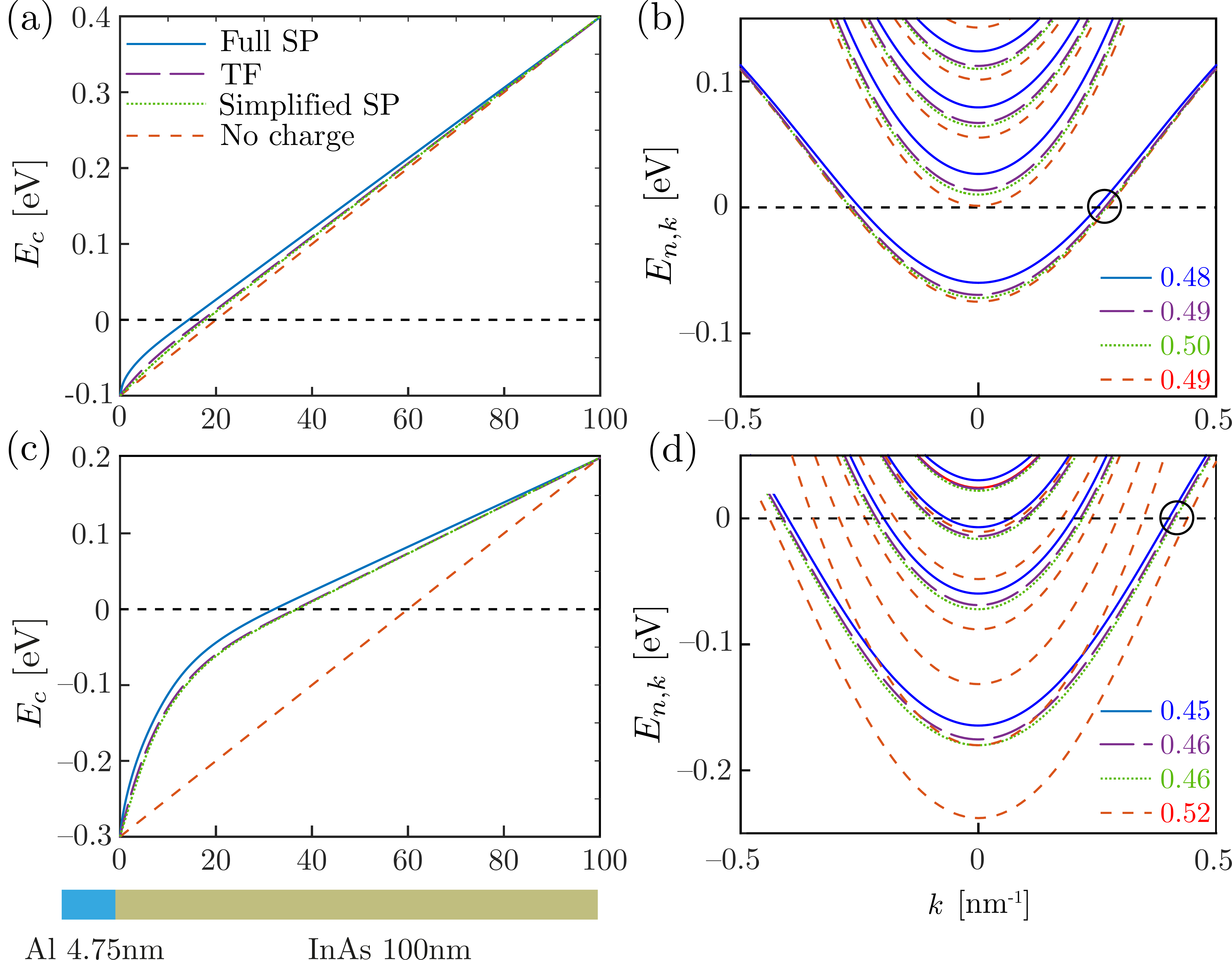}
\caption{(Color online) Self-consistent band edge profiles and band structures obtained using the methods described in Appendix~\ref{sec:BandStructureComparison} with $L_{1}=4.75\,\rm{nm}$ and $V_{\D}=-0.5\,\V$. Results for both $\Phi=0.1\rm{eV}$ and $\Phi=0.3\rm{eV}$ are shown. (a) Band edge profiles obtained for $\Phi=0.1\,\rm{eV}$. (b) Band structures obtained for $\Phi=0.1\,\rm{eV}$. The numbers in the lower left corner are the weights at the Fermi energy of the bands marked by the black circle. (c) Band edge profiles obtained for $\Phi=0.3\,\rm{eV}$. (d) Band structures obtained for $\Phi=0.3\,\rm{eV}$. The numbers in the lower left corner are the weights at the Fermi energy of the bands marked by the black circle.}
\label{fig:AppendixFigure1}
\end{figure*}

\subsection{The Schr\"odinger equation}

The Schr\"odinger equation \eqref{eq:Schrodinger} is solved on a 1D grid using the finite difference approximation:
\begin{widetext}
\begin{align}
\frac{-\hbar^2}{z_{i+1}-z_{i-1}}  \left( \frac{1}{m^{*}_{i+1/2}}\frac{\psi_{i+1}-\psi_{i}}{z_{i+1}-z_{i}}-\frac{1}{m^{*}_{i-1/2}}\frac{\psi_{i}-\psi_{i-1}}{z_{i}-z_{i-1}}\right)    + \Bigl(\frac{\hbar^2 k^2}{2m_{i}^{*}}+E_{c,i}\Bigr)\psi_{i} = E\psi_{i}.  \label{eq:SchrodingerNumeric}
\end{align}
\end{widetext}
Here $m^{*}_{i+1/2}$ denotes the average value of the effective mass on the two grid points i and i+1. Since the Fermi wave length of the metal is orders of magnitude smaller than that of the semiconductor it is advantageous to make the discretization more coarse in the semiconductor region. The results presented in this work were obtained using a fixed grid spacing of 0.1\,\AA \ in the Al region and 2\,\AA \ in the InAs region. To obtain the solutions to the system of Eq.~\eqref{eq:SchrodingerNumeric}, we solve it as an eigenvalue equation. We enforce hard-wall boundary condition by setting the wave functions to zero at the ends of the 1D grid.

\subsection{Obtaining the self-consistent solution}

Our Schr\"odinger-Poisson approach relies on self-consistently solving Eqs.~\eqref{eq:Poisson},~\eqref{eq:Schrodinger} and~\eqref{eq:ChargeDensitySP}. For this we employ a simple mixing scheme, where the input electrostatic potential used in each iteration is a simple mixing of the input and output electrostatic potential of the previous iteration
\begin{equation}\label{eq:phii}
\phi_{\rm{in}}^{i}(z) = \kappa \phi_{\rm{out}}^{i-1}(z) + (1-\kappa) \phi_{\rm{in}}^{i-1}(z).
\end{equation}
In our calculations we used $\kappa=0.1$. For the initial input, we use $\phi_{\rm{in}}^{1}(z)=V_{\D}z/L_{2}$. While the authors of Ref.~\cite{Vuik2016} have shown that more sophisticated mixing schemes such as Anderson mixing leads to a faster convergence, we find that the simple mixing scheme above provides reasonably fast convergence within the first 50-100 self-consistent iterations.

\section{Influence of band bending on the hybridization}\label{sec:BandStructureComparison}

The hybrid band structures shown in Sec.~\ref{sec:BS} were obtained using $E_{c}(z)$ obtained from the Schr\"odinger-Poisson approach. This procedure is computationally demanding since it requires solving Eq.~\eqref{eq:Schrodinger} a larger number of times within each self-consistent iteration when calculating the electronic density from Eq.~\eqref{eq:ChargeDensitySP}. In this section, we explore the consequences of employing simpler and computationally faster approaches for calculating $E_{c}(z)$. Specifically, we compare the results of Sec.~\ref{sec:BS} to those obtained when $E_{c}(z)$ is determined by:

\begin{enumerate}
\item
The Thomas-Fermi approach described in Sec.~\ref{sec:TF}.

\item A simplified Schr\"odinger-Poisson approach, where Eq.~\eqref{eq:Schrodinger} is solved only in the InAs region with boundary conditions $\psi_{n}(0)=\psi_{n}(L_{2})=0$. In this case the wave functions are independent of $k$ and the density in the semiconductor region is found by integrating over the 2D density of states for each subband yielding
\begin{equation}\label{rhoz}
\rho(z) =- \sum_{n} \frac{m_\InAs\lvert E_{n} \lvert}{\pi^2 \hbar^2} \lvert \psi_{n}(z) \lvert^2\Theta(-E_n).
\end{equation}

\item With $\rho(z)=0$ in the semiconductor, i.e. neglecting band bending due to charge in the InAs such that
\begin{equation}\label{Ecz}
E_{c}(z) = -\Phi -eV_{\D} z/L_{2}.
\end{equation}
\end{enumerate}

Our results are summarized in Fig.~\ref{fig:AppendixFigure1} where we show both the resulting band bending and the band structure plots from the different approaches for $\Phi=0.1\,\rm{eV}$ and $\Phi=0.3\,\rm{eV}$. We have here chosen parameters such that a strongly hybridized band crosses the Fermi level, and the corresponding InAs weights at the Fermi energy (indicated by the circle) are shown in the bottom left corners of Figs.~\ref{fig:AppendixFigure1}(b) and~(d).

When $\Phi=0.1\,\rm{eV}$ (Figs.~\ref{fig:AppendixFigure1}(a,b)) the density in the InAs region is low, and the band edges exhibit only a slight bending, staying close to the constant slope found without density in the semiconductor. Notably the strongest bending is found when the full Schr\"odinger-Poisson approach is employed. This is due to the hybridization with the Al which induces a large electron density close to the Al-InAs interface (see Fig.~\ref{fig:DeviceSketch}(d)). In the case $\Phi=0.3\,\rm{eV}$, the band bending profiles are much stronger and deviate substantially from the solution without charge. Thus, the strongest bending is found from the full Schr\"odinger-Poisson approach, but both Thomas-Fermi and simplified Schr\"odinger-Poisson methods yield similar results.

The same conclusion holds for the band structure plot of Fig.~\ref{fig:AppendixFigure1}(d). Here the band structures obtained with band bending are reasonably similar, while the one obtained with constant slope evidently contains additional bands below the Fermi level due to the more shallow band profile. Interestingly, it appears that the weight at the Fermi energy of the strongly hybridized band is only weakly dependent on the exact band bending profile and all approaches yield comparatively similar results.

\section{Analytical approach to hybridization}\label{sec:Analytix}

\begin{figure*}[t!]
\includegraphics[width=\textwidth]{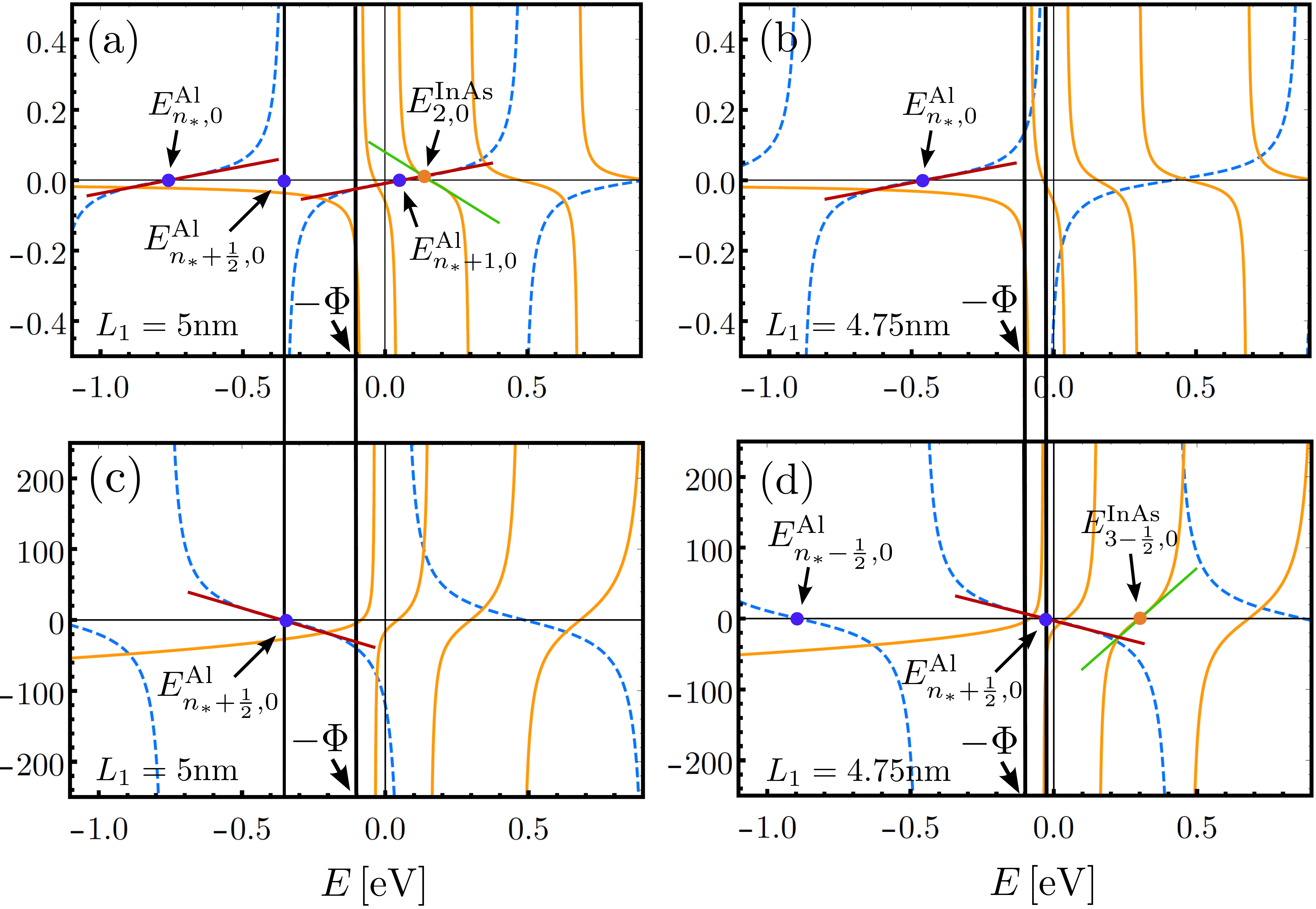}
\caption{Plot of the l.h.s. (dashed blue) and r.h.s. (solid gold) of the direct and inverted transcedental Eqs.~\eqref{eq:transcendental} and~\eqref{eq:transcendentalInverted} for $\Phi=0.1\,\eV$ and $L_2'=16\,\nm$. In (a,c) $L_1=5\,\nm$ and in (b,d) $L_1=4.75\,\nm$. The eigenenergies are obtained when the l.h.s. and r.h.s. lines cross. Depending on the energy regime it is convenient to use the direct or inverted transcedental equation in order to obtain the eigenspectrum. The approximate analytical expressions for the eigenenergies are obtained via a Taylor expansion of the l.h.s. or/and the r.h.s. about zeros of the respective tangent or cotangent. These zeros are related to the pure Al and InAs levels and the values $E_{n,k}^\Al$ and $E_{n,k}^\InAs$, as well as $E_{n+1/2,k}^\Al$ and $E_{n+1/2,k}^\InAs$. (a) is employed for inferring weakly modified Al-like bands and InAs-like bands within scenario~\textbf{I}. (b) is employed for inferring weakly modified Al-like bands within scenario~\textbf{II}. (c) is employed for inferring the Al-InAs gluing segments within scenario~\textbf{I}. (d) is employed for inferring modified InAs-like bands within scenario~\textbf{II}.}
\label{fig:GraphicalSols}
\end{figure*}

In this appendix, we demonstrate how to obtain approximate analytical expressions for the band structure pro\-per\-ties of the square-well model of Sec.~\ref{sec:RecModel}. We start from Eq.~\eqref{matchcondition} and rewrite the transcendental equation as
\begin{equation}
\frac{\tan\left(k_\Al L_1\right)}{\frac{k_\Al L_1}{\pi}}=-\frac{m_\InAs L_2'}{m_\Al L_1}\frac{\tan\left(k_\InAs L_2'\right)}{\frac{k_\InAs L_2'}{\pi}},\label{eq:transcendental}
\end{equation}
or bring it to its inverted form
\begin{align}
\frac{k_\Al L_1}{\pi}\cot\left(k_\Al L_1\right)=-\frac{m_\Al L_1}{m_\InAs L_2'}\frac{k_\InAs L_2'}{\pi}\cot\left(k_\InAs L_2'\right).\label{eq:transcendentalInverted}
\end{align}

Fig.~\ref{fig:GraphicalSols} depicts the functions of the l.h.s. (dashed blue) and r.h.s. (solid gold) of Eqs.~\eqref{eq:transcendental} and~\eqref{eq:transcendentalInverted} for $\Phi=0.1\,\eV$, $L_2'=16\,\nm$, $k=0$, and Al width $L_1=5\,\nm$ and $L_1=4.75\,\nm$, respectively. The energy eigenstates of the hybrid band structure are given by the crossing points of the two functions. For low energies we have essentially solutions corresponding to isolated Al. Instead, for energies in the vicinity of the pure InAs levels, we find solutions emerging from the hybridization of InAs and Al. In Fig.~\ref{fig:GraphicalSols} the top and corresponding bottom panels, i.e. (a,c) and (b,d), lead to identical solutions. Nevertheless, we have included them both since, de\-pen\-ding on the energy regime, it is more convenient to employ the inverted~\eqref{eq:transcendentalInverted} instead of the direct transcedental equation~\eqref{eq:transcendental}.

The aim is to Taylor expand the l.h.s. or/and r.h.s., of the respective transcedental equation employed, about zero. Depending on the case, one performs a linear or quadratic Taylor expansion. If we work with Eq.~\eqref{eq:transcendental} we can expand the l.h.s. as follows
\begin{align}
\frac{\tan\left(k_\Al L_1\right)}{\frac{k_\Al L_1}{\pi}}\approx
\frac{\pi}{2}\frac{\delta E}{E_{n,k=0}^{\rm Al}+E_F}\left[1-\frac{3}{4}\frac{\delta E}{E_{n,k=0}^{\rm Al}+E_F}\right]
,\label{eq:Approximation1}
\end{align}
with the energy shift $\delta E=E-E_{n,k}^\Al$ being measured from a pure Al level, which yields a zero tangent since $k_\Al L_1/\pi=n\in\mathbb{N}^+$. A similar method is followed if we want to expand the r.h.s. of the same equation about the pure InAs levels.

If it is instead preferable to employ Eq.~\eqref{eq:transcendentalInverted}, then one should expand about a zero of the respective l.h.s. or r.h.s.. For instance, if we wish to expand the l.h.s. about a zero of the cotangent obtained for
\begin{equation}
\frac{k_\Al L_1}{\pi}=n+\frac{1}{2}\quad{\rm with}\quad n\in\mathbb{N}^+,
\end{equation}
we have
\begin{eqnarray}
&&\frac{k_\Al L_1}{\pi}\cot\left(k_\Al L_1\right)\approx \nonumber\\
&&-\frac{\pi}{2}\frac{\delta E}{E_{1,k=0}^{\rm Al}+E_F}\left[1+\frac{1}{4}\frac{\delta E}{E_{n+\frac{1}{2},k=0}^{\rm Al}+E_F}\right].\\\nonumber\label{eq:Approximation2}
\end{eqnarray}
Here the energy shift $\delta E=E-E_{n+\frac{1}{2},k}^\Al$ is measured re\-la\-ti\-ve to the $n+\frac{1}{2}$th pure Al level. In reality there is no such a pure Al level before contact with InAs, but this notation is convenient because it reflects that we are focusing on energy eigenstates which appear due to the Al-InAs hybridization. One can further expand the r.h.s. about the $s+\frac{1}{2}$th pure InAs level, with $s\in\mathbb{N}^+$, in a similar fashion.

Before proceeding with obtaining the various appro\-xi\-mate analytical expression discussed in the text, let us remark that our approach generally holds for the case of small Al widths where the spacing of the pure Al levels is much larger that the spacing of the pure InAs levels. Our approximation further holds for arbitrary values of $L_2'$. However, it can modify the number of Al-InAs bands for which our method is valid.

\subsection{Solutions with $\bm{k_\Al\in\mathbb{R}}$ and $\bm{k_\InAs\equiv i|k_\InAs|\in\mathbb{I}}$}

After applying the matching conditions, we find that the wave function for such a solution with $E_{n,k}$ has the approximate form
\begin{equation}
\psi_{n,k}(z)\approx (-1)^{n+1}\sqrt{w_{\rm Al}}\sqrt{\frac{2}{L_1}}\sin\left[k_\Al(z+L_1)\right],
\end{equation}
for $z\in[-L_1,0]$ and
\begin{equation}
\psi_{n,k}(z)\approx\sqrt{w_{\rm InAs}}\sqrt{\frac{2}{\xi_\InAs}}\frac{\sinh\left[|k_\InAs|(L_2'-z)\right]}{\sinh(|k_\InAs|L_2')},
\end{equation}
for $z\in[0,L_2']$. The InAs and Al weights are defined as
\begin{equation}
w_{\rm InAs}=\frac{\int_{0}^{L_2'}dz\ph|\psi_{n,k}(z)|^2}{\int_{-L_1}^{L_2'}dz\ph|\psi_{n,k}(z)|^2}\phd{\rm and}\phd w_\Al=1-w_\InAs.
\end{equation}

We first discuss the weakly modified pure Al levels, with $n\leq n_*$ for all $k$ and $n>n_*$ for $k>k_h$. In this case we start from Eq.~\ref{eq:transcendental}. We linearize the l.h.s. about $E_{n,k}^\Al$, consider $\tanh\left(|k_\InAs| L_2'\right)\approx1$, and set $E=E_{n,k}^\Al$ on the r.h.s.. In Figs.~\ref{fig:GraphicalSols}(a,b) we show details for the $n=n_*$ level. We find that the eigenenergies appro\-xi\-ma\-te\-ly read
\begin{eqnarray}
&&E_{n,k}=E_{n,k}^\Al\\
&&-\frac{2}{\pi}\frac{m_\InAs L_2'}{m_\Al L_1}
\sqrt{\frac{E_{1,k=0}^\InAs+\Phi}{\hbar^2 k^2/(2m_\InAs)-\Phi-E_{n,k}^{\Al}}}\left(E_{n,k=0}^\Al+E_F\right).\nonumber
\end{eqnarray}
Using the above, we obtained Eqs.~\eqref{eq:xiAl} and~\eqref{eq:WAl}.

We proceed with the band segments for $n>n_*$ and $k\in[k_\ell,k_h]$. In this case we consider the inverted transcedental Eq.~\eqref{eq:transcendentalInverted}. We linearize the l.h.s. about $E_{n-\frac{1}{2},k}^\Al$, consider $\coth\left(|k_\InAs| L_2'\right)\approx1$, and set $E=E_{n-\frac{1}{2},k}^\Al$ on the r.h.s.. See also Fig.~\eqref{fig:GraphicalSols}(c). We find the eigenenergies
\begin{widetext}
\begin{equation}
E_{n,k}=\frac{\hbar^2 k^2}{2m_\InAs}-
\left[\sqrt{\frac{\hbar^2k^2/(2m_\InAs)-\Phi-E_{n-\frac{1}{2},k}^{\rm Al}}{E_{1,k=0}^{\rm Al}+E_F}+\left(\frac{1}{\pi}\sqrt{\frac{m_\Al}{m_\InAs}}\right)^2}-\frac{1}{\pi}\sqrt{\frac{m_\Al}{m_\InAs}}\right]^2\left(E_{1,k=0}^{\rm Al}+E_F\right).
\end{equation}
\end{widetext}

Note that the above expression does not depend on $L_2'$ at this level of approximation. Using the above, we obtained Eqs.~\eqref{eq:Evawinas} and~\eqref{winas}.

\subsection{Solutions with $\bm{k_\Al\in\mathbb{R}}$ and $\bm{k_\InAs\in\mathbb{R}}$}

After applying the matching conditions, we find that the wave function for such a solution with $E_{n,k}$ has the approximate form
\begin{equation}
\psi_{n,k}(z)\approx (-1)^{n+1}\sqrt{w_{\rm Al}}\sqrt{\frac{2}{L_1}}\sin\left[k_\Al(z+L_1)\right].
\end{equation}
for $z\in[-L_1,0]$ and
\begin{equation}
\psi_{n,k}(z)\approx \sqrt{w_{\rm InAs}}\sqrt{\frac{2}{L_2'}}\sin\left[k_\InAs(L_2'-z)\right],
\end{equation}
for $z\in[0,L_2']$ with the weights discussed in the main text. These solutions appear for $k<k_\ell$ defined in Sec.~\ref{sec:Analytics}. One distinguishes two scenarios \textbf{I} and \textbf{II}.

\textbf{Scenario I} In this case we use Eq.~\eqref{eq:transcendental} and linearize both sides. We linearize the l.h.s. about the $E_{n_*+1,k}^\Al$ level and the r.h.s. about the $E_{n-n_*-1,k}^\InAs$ level. See Fig.~\ref{fig:GraphicalSols}(a) for $n=n_*+2$. This approximation led to Eqs.~\eqref{EInAsmodified} and~\eqref{westimate}, and holds as long as the linear approximation of the l.h.s. is valid, i.e. $E_{n-n_*-1,k}^\InAs  \lesssim E_{n_*+\frac{5}{4},k}^\Al$.

\textbf{Scenario II} In this case we use Eq.~\eqref{eq:transcendentalInverted} and li\-nea\-ri\-ze both sides. We linearize the l.h.s. about the $E_{n_*+\frac{1}{2},k}^\Al$ level and the r.h.s. about the $E_{n-n_*-\frac{1}{2},k}^\InAs$ level. See Fig.~\ref{fig:GraphicalSols}(d) for $n=n_*+3$. This approximation led to Eqs.~\eqref{EInAsmodified2} and~\eqref{westimate2}. Note that this approximation holds as long as the linear approximation of the l.h.s. is valid, i.e. $E_{n-n_*-\frac{1}{2},k}^\InAs\lesssim E_{n_*+\frac{3}{4},k}^\Al$. In this case, for $n=n_*+3$ we are on the borderline of our approximation's validity. For larger $L_2'$, e.g. $L_2'\sim80\nm$ we have to expand up to quadratic order the r.h.s. in order to obtain a good approximate solution, and in this case the energy reads
\begin{widetext}
\begin{align}
E_{n,k}=E^{\rm InAs}_{n-n_*-\frac{1}{2},k}+2\left\{
\sqrt{\left(1+\frac{L_1}{L_2'}\right)^2+\frac{L_1}{L_2'}
\frac{E^\Al_{n_*+\frac{1}{2},k}-E^\InAs_{n-n_*-\frac{1}{2},k}}{E^\InAs_{n-n_*-\frac{1}{2},k=0}+\Phi}
}-\left(1+\frac{L_1}{L_2'}\right)\right\}\left(E^\InAs_{n-n_*-\frac{1}{2},k=0}+\Phi\right),
\end{align}
\end{widetext}
which is applicable for bands satisfying $E^\Al_{n_*+1/2,k}>E^\InAs_{n-n_*-1/2,k}$. The approach can be extended to higher energies by separating the energy interval in regions where the direct or inverted transcendental equation is best.


%\bibliographystyle{apsrev4-1}
%\bibliography{litlist,../../../../Majorana/Majorana}

\begin{thebibliography}{45}%
\makeatletter
\providecommand \@ifxundefined [1]{%
 \@ifx{#1\undefined}
}%
\providecommand \@ifnum [1]{%
 \ifnum #1\expandafter \@firstoftwo
 \else \expandafter \@secondoftwo
 \fi
}%
\providecommand \@ifx [1]{%
 \ifx #1\expandafter \@firstoftwo
 \else \expandafter \@secondoftwo
 \fi
}%
\providecommand \natexlab [1]{#1}%
\providecommand \enquote  [1]{``#1''}%
\providecommand \bibnamefont  [1]{#1}%
\providecommand \bibfnamefont [1]{#1}%
\providecommand \citenamefont [1]{#1}%
\providecommand \href@noop [0]{\@secondoftwo}%
\providecommand \href [0]{\begingroup \@sanitize@url \@href}%
\providecommand \@href[1]{\@@startlink{#1}\@@href}%
\providecommand \@@href[1]{\endgroup#1\@@endlink}%
\providecommand \@sanitize@url [0]{\catcode `\\12\catcode `\$12\catcode
  `\&12\catcode `\#12\catcode `\^12\catcode `\_12\catcode `\%12\relax}%
\providecommand \@@startlink[1]{}%
\providecommand \@@endlink[0]{}%
\providecommand \url  [0]{\begingroup\@sanitize@url \@url }%
\providecommand \@url [1]{\endgroup\@href {#1}{\urlprefix }}%
\providecommand \urlprefix  [0]{URL }%
\providecommand \Eprint [0]{\href }%
\providecommand \doibase [0]{http://dx.doi.org/}%
\providecommand \selectlanguage [0]{\@gobble}%
\providecommand \bibinfo  [0]{\@secondoftwo}%
\providecommand \bibfield  [0]{\@secondoftwo}%
\providecommand \translation [1]{[#1]}%
\providecommand \BibitemOpen [0]{}%
\providecommand \bibitemStop [0]{}%
\providecommand \bibitemNoStop [0]{.\EOS\space}%
\providecommand \EOS [0]{\spacefactor3000\relax}%
\providecommand \BibitemShut  [1]{\csname bibitem#1\endcsname}%
\let\auto@bib@innerbib\@empty
%</preamble>
\bibitem [{Sze()}]{SzeNgBook}%
  \BibitemOpen
  \href@noop {} {}\bibinfo {note} {S. M. Sze and K. K. Ng, Physics of
  Semiconductor Devices, 3rd Edition, (Wiley-Interscience, New Jersey,
  2006).}\BibitemShut {Stop}%
\bibitem [{\citenamefont {Bardeen}(1947)}]{Bardeen1947}%
  \BibitemOpen
  \bibfield  {author} {\bibinfo {author} {\bibfnamefont {J.}~\bibnamefont
  {Bardeen}},\ }\bibfield  {title} {\emph {\bibinfo {title} {{Surface states
  and rectification at a metal semi-conductor contact}},\ }}\href {\doibase
  10.1103/PhysRev.71.717} {\bibfield  {journal} {\bibinfo  {journal}
  {Phys.~Rev.}\ }\textbf {\bibinfo {volume} {71}},\ \bibinfo {pages} {717}
  (\bibinfo {year} {1947})}\BibitemShut {NoStop}%
\bibitem [{\citenamefont {Heine}(1965)}]{Heine1965}%
  \BibitemOpen
  \bibfield  {author} {\bibinfo {author} {\bibfnamefont {V.}~\bibnamefont
  {Heine}},\ }\bibfield  {title} {\emph {\bibinfo {title} {{Theory of Surface
  States}},\ }}\href {\doibase 10.1103/PhysRev.138.A1689} {\bibfield  {journal}
  {\bibinfo  {journal} {Phys.~Rev.}\ }\textbf {\bibinfo {volume} {138}},\
  \bibinfo {pages} {A1689} (\bibinfo {year} {1965})}\BibitemShut {NoStop}%
\bibitem [{\citenamefont {Beenakker}(2013)}]{BeenakkerReview}%
  \BibitemOpen
  \bibfield  {author} {\bibinfo {author} {\bibfnamefont {C.~W.~J.}\
  \bibnamefont {Beenakker}},\ }\bibfield  {title} {\emph {\bibinfo {title}
  {{Search for Majorana fermions in superconductors}},\ }}\href {\doibase
  10.1146/annurev-conmatphys-030212-184337} {\bibfield  {journal} {\bibinfo
  {journal} {Annu. Rev. Condens. Matter Phys.}\ }\textbf {\bibinfo {volume}
  {4}},\ \bibinfo {pages} {113} (\bibinfo {year} {2013})}\BibitemShut {NoStop}%
\bibitem [{\citenamefont {Alicea}(2012)}]{AliceaReview}%
  \BibitemOpen
  \bibfield  {author} {\bibinfo {author} {\bibfnamefont {J.}~\bibnamefont
  {Alicea}},\ }\bibfield  {title} {\emph {\bibinfo {title} {{New directions in
  the pursuit of Majorana fermions in solid state systems.}}\ }}\href {\doibase
  10.1088/0034-4885/75/7/076501} {\bibfield  {journal} {\bibinfo  {journal}
  {Rep. Prog. Phys.}\ }\textbf {\bibinfo {volume} {75}},\ \bibinfo {pages}
  {076501} (\bibinfo {year} {2012})}\BibitemShut {NoStop}%
\bibitem [{\citenamefont {Leijnse}\ and\ \citenamefont
  {Flensberg}(2012)}]{LeijnseReview}%
  \BibitemOpen
  \bibfield  {author} {\bibinfo {author} {\bibfnamefont {M.}~\bibnamefont
  {Leijnse}}\ and\ \bibinfo {author} {\bibfnamefont {K.}~\bibnamefont
  {Flensberg}},\ }\bibfield  {title} {\emph {\bibinfo {title} {{Introduction to
  topological superconductivity and Majorana fermions}},\ }}\href {\doibase
  10.1088/0268-1242/27/12/124003} {\bibfield  {journal} {\bibinfo  {journal}
  {Semicond.~Sci.~Technol.}\ }\textbf {\bibinfo {volume} {27}},\ \bibinfo
  {pages} {124003} (\bibinfo {year} {2012})}\BibitemShut {NoStop}%
\bibitem [{\citenamefont {Aguado}(2017)}]{AguadoReview}%
  \BibitemOpen
  \bibfield  {author} {\bibinfo {author} {\bibfnamefont {R.}~\bibnamefont
  {Aguado}},\ }\href {\doibase 10.1393/ncr/i2017-10141-9} {\bibfield  {journal}
  {\bibinfo  {journal} {La Rivista del Nuovo Cimento}\ }\textbf {\bibinfo
  {volume} {40}},\ \bibinfo {pages} {523} (\bibinfo {year} {2017})}\BibitemShut
  {NoStop}%
\bibitem [{\citenamefont {Lutchyn}\ \emph {et~al.}(2018)\citenamefont
  {Lutchyn}, \citenamefont {Bakkers}, \citenamefont {Kouwenhoven},
  \citenamefont {Krogstrup}, \citenamefont {Marcus},\ and\ \citenamefont
  {Oreg}}]{LutchynReview}%
  \BibitemOpen
  \bibfield  {author} {\bibinfo {author} {\bibfnamefont {R.~M.}\ \bibnamefont
  {Lutchyn}}, \bibinfo {author} {\bibfnamefont {E.~P. A.~M.}\ \bibnamefont
  {Bakkers}}, \bibinfo {author} {\bibfnamefont {L.~P.}\ \bibnamefont
  {Kouwenhoven}}, \bibinfo {author} {\bibfnamefont {P.}~\bibnamefont
  {Krogstrup}}, \bibinfo {author} {\bibfnamefont {C.~M.}\ \bibnamefont
  {Marcus}}, \ and\ \bibinfo {author} {\bibfnamefont {Y.}~\bibnamefont
  {Oreg}},\ }\href@noop {} {\bibfield  {journal} {\bibinfo  {journal} {Nature
  Review Materials}\ }\textbf {\bibinfo {volume} {3}},\ \bibinfo {pages} {52}
  (\bibinfo {year} {2018})}\BibitemShut {NoStop}%
\bibitem [{\citenamefont {Kitaev}(2001)}]{Kitaev2001}%
  \BibitemOpen
  \bibfield  {author} {\bibinfo {author} {\bibfnamefont {A.~Y.}\ \bibnamefont
  {Kitaev}},\ }\bibfield  {title} {\emph {\bibinfo {title} {{Unpaired Majorana
  fermions in quantum wires}},\ }}\href {\doibase 10.1070/1063-7869/44/10S/S29}
  {\bibfield  {journal} {\bibinfo  {journal} {Phys.~Usp.}\ }\textbf {\bibinfo
  {volume} {44}},\ \bibinfo {pages} {131} (\bibinfo {year} {2001})}\BibitemShut
  {NoStop}%
\bibitem [{\citenamefont {Ivanov}(2001)}]{Ivanov2001}%
  \BibitemOpen
  \bibfield  {author} {\bibinfo {author} {\bibfnamefont {D.~A.}\ \bibnamefont
  {Ivanov}},\ }\bibfield  {title} {\emph {\bibinfo {title} {{Non-Abelian
  Statistics of Half-Quantum Vortices in p-Wave Superconductors}},\ }}\href
  {\doibase 10.1103/PhysRevLett.86.268} {\bibfield  {journal} {\bibinfo
  {journal} {Phys.~Rev.~Lett.}\ }\textbf {\bibinfo {volume} {86}},\ \bibinfo
  {pages} {268} (\bibinfo {year} {2001})}\BibitemShut {NoStop}%
\bibitem [{\citenamefont {Alicea}\ \emph {et~al.}(2011)\citenamefont {Alicea},
  \citenamefont {Oreg}, \citenamefont {Refael}, \citenamefont {von Oppen},\
  and\ \citenamefont {Fisher}}]{Alicea2011}%
  \BibitemOpen
  \bibfield  {author} {\bibinfo {author} {\bibfnamefont {J.}~\bibnamefont
  {Alicea}}, \bibinfo {author} {\bibfnamefont {Y.}~\bibnamefont {Oreg}},
  \bibinfo {author} {\bibfnamefont {G.}~\bibnamefont {Refael}}, \bibinfo
  {author} {\bibfnamefont {F.}~\bibnamefont {von Oppen}}, \ and\ \bibinfo
  {author} {\bibfnamefont {M.~P.~A.}\ \bibnamefont {Fisher}},\ }\bibfield
  {title} {\emph {\bibinfo {title} {{Non-Abelian statistics and topological
  quantum information processing in 1D wire networks}},\ }}\href {\doibase
  10.1038/nphys1915} {\bibfield  {journal} {\bibinfo  {journal} {Nat.~Phys.}\
  }\textbf {\bibinfo {volume} {7}},\ \bibinfo {pages} {412} (\bibinfo {year}
  {2011})}\BibitemShut {NoStop}%
\bibitem [{\citenamefont {Aasen}\ \emph {et~al.}(2016)\citenamefont {Aasen},
  \citenamefont {Hell}, \citenamefont {Mishmash}, \citenamefont {Higginbotham},
  \citenamefont {Danon}, \citenamefont {Leijnse}, \citenamefont {Jespersen},
  \citenamefont {Folk}, \citenamefont {Marcus}, \citenamefont {Flensberg},\
  and\ \citenamefont {Alicea}}]{Aasen2016}%
  \BibitemOpen
  \bibfield  {author} {\bibinfo {author} {\bibfnamefont {D.}~\bibnamefont
  {Aasen}}, \bibinfo {author} {\bibfnamefont {M.}~\bibnamefont {Hell}},
  \bibinfo {author} {\bibfnamefont {R.~V.}\ \bibnamefont {Mishmash}}, \bibinfo
  {author} {\bibfnamefont {A.}~\bibnamefont {Higginbotham}}, \bibinfo {author}
  {\bibfnamefont {J.}~\bibnamefont {Danon}}, \bibinfo {author} {\bibfnamefont
  {M.}~\bibnamefont {Leijnse}}, \bibinfo {author} {\bibfnamefont {T.~S.}\
  \bibnamefont {Jespersen}}, \bibinfo {author} {\bibfnamefont {J.~A.}\
  \bibnamefont {Folk}}, \bibinfo {author} {\bibfnamefont {C.~M.}\ \bibnamefont
  {Marcus}}, \bibinfo {author} {\bibfnamefont {K.}~\bibnamefont {Flensberg}}, \
  and\ \bibinfo {author} {\bibfnamefont {J.}~\bibnamefont {Alicea}},\
  }\bibfield  {title} {\emph {\bibinfo {title} {{Milestones Toward
  Majorana-Based Quantum Computing}},\ }}\href {\doibase
  10.1103/PhysRevX.6.031016} {\bibfield  {journal} {\bibinfo  {journal}
  {Phys.~Rev.~X}\ }\textbf {\bibinfo {volume} {6}},\ \bibinfo {pages} {031016}
  (\bibinfo {year} {2016})}\BibitemShut {NoStop}%
\bibitem [{\citenamefont {Plugge}\ \emph {et~al.}(2017)\citenamefont {Plugge},
  \citenamefont {Rasmussen}, \citenamefont {Egger},\ and\ \citenamefont
  {Flensberg}}]{Plugge2017}%
  \BibitemOpen
  \bibfield  {author} {\bibinfo {author} {\bibfnamefont {S.}~\bibnamefont
  {Plugge}}, \bibinfo {author} {\bibfnamefont {A.}~\bibnamefont {Rasmussen}},
  \bibinfo {author} {\bibfnamefont {R.}~\bibnamefont {Egger}}, \ and\ \bibinfo
  {author} {\bibfnamefont {K.}~\bibnamefont {Flensberg}},\ }\bibfield  {title}
  {\emph {\bibinfo {title} {Majorana box qubits},\ }}\href {\doibase
  10.1088/1367-2630/aa54e1} {\bibfield  {journal} {\bibinfo  {journal} {New
  Journal of Physics}\ }\textbf {\bibinfo {volume} {19}},\ \bibinfo {pages}
  {012001} (\bibinfo {year} {2017})}\BibitemShut {NoStop}%
\bibitem [{\citenamefont {Karzig}\ \emph {et~al.}(2017)\citenamefont {Karzig},
  \citenamefont {Knapp}, \citenamefont {Lutchyn}, \citenamefont {Bonderson},
  \citenamefont {Hastings}, \citenamefont {Nayak}, \citenamefont {Alicea},
  \citenamefont {Flensberg}, \citenamefont {Plugge}, \citenamefont {Oreg},
  \citenamefont {Marcus},\ and\ \citenamefont {Freedman}}]{Karzig2017}%
  \BibitemOpen
  \bibfield  {author} {\bibinfo {author} {\bibfnamefont {T.}~\bibnamefont
  {Karzig}}, \bibinfo {author} {\bibfnamefont {C.}~\bibnamefont {Knapp}},
  \bibinfo {author} {\bibfnamefont {R.~M.}\ \bibnamefont {Lutchyn}}, \bibinfo
  {author} {\bibfnamefont {P.}~\bibnamefont {Bonderson}}, \bibinfo {author}
  {\bibfnamefont {M.~B.}\ \bibnamefont {Hastings}}, \bibinfo {author}
  {\bibfnamefont {C.}~\bibnamefont {Nayak}}, \bibinfo {author} {\bibfnamefont
  {J.}~\bibnamefont {Alicea}}, \bibinfo {author} {\bibfnamefont
  {K.}~\bibnamefont {Flensberg}}, \bibinfo {author} {\bibfnamefont
  {S.}~\bibnamefont {Plugge}}, \bibinfo {author} {\bibfnamefont
  {Y.}~\bibnamefont {Oreg}}, \bibinfo {author} {\bibfnamefont {C.~M.}\
  \bibnamefont {Marcus}}, \ and\ \bibinfo {author} {\bibfnamefont {M.~H.}\
  \bibnamefont {Freedman}},\ }\bibfield  {title} {\emph {\bibinfo {title}
  {{Scalable designs for quasiparticle-poisoning-protected topological quantum
  computation with Majorana zero modes}},\ }}\href {\doibase
  10.1103/PhysRevB.95.235305} {\bibfield  {journal} {\bibinfo  {journal}
  {Phys.~Rev.~B}\ }\textbf {\bibinfo {volume} {95}},\ \bibinfo {pages} {235305}
  (\bibinfo {year} {2017})}\BibitemShut {NoStop}%
\bibitem [{\citenamefont {Mourik}\ \emph {et~al.}(2012)\citenamefont {Mourik},
  \citenamefont {Zuo}, \citenamefont {Frolov}, \citenamefont {Plissard},
  \citenamefont {Bakkers},\ and\ \citenamefont {Kouwenhoven}}]{Mourik2012}%
  \BibitemOpen
  \bibfield  {author} {\bibinfo {author} {\bibfnamefont {V.}~\bibnamefont
  {Mourik}}, \bibinfo {author} {\bibfnamefont {K.}~\bibnamefont {Zuo}},
  \bibinfo {author} {\bibfnamefont {S.~M.}\ \bibnamefont {Frolov}}, \bibinfo
  {author} {\bibfnamefont {S.~R.}\ \bibnamefont {Plissard}}, \bibinfo {author}
  {\bibfnamefont {E.~P. A.~M.}\ \bibnamefont {Bakkers}}, \ and\ \bibinfo
  {author} {\bibfnamefont {L.~P.}\ \bibnamefont {Kouwenhoven}},\ }\bibfield
  {title} {\emph {\bibinfo {title} {{Signatures of Majorana fermions in hybrid
  superconductor-semiconductor nanowire devices.}}\ }}\href {\doibase
  10.1126/science.1222360} {\bibfield  {journal} {\bibinfo  {journal}
  {Science}\ }\textbf {\bibinfo {volume} {336}},\ \bibinfo {pages} {1003}
  (\bibinfo {year} {2012})}\BibitemShut {NoStop}%
\bibitem [{\citenamefont {Krogstrup}\ \emph {et~al.}(2015)\citenamefont
  {Krogstrup}, \citenamefont {Ziino}, \citenamefont {Chang}, \citenamefont
  {Albrecht}, \citenamefont {Madsen}, \citenamefont {Johnson}, \citenamefont
  {Nyg{\aa}rd}, \citenamefont {Marcus},\ and\ \citenamefont
  {Jespersen}}]{Krogstrup2015}%
  \BibitemOpen
  \bibfield  {author} {\bibinfo {author} {\bibfnamefont {P.}~\bibnamefont
  {Krogstrup}}, \bibinfo {author} {\bibfnamefont {N.~L.~B.}\ \bibnamefont
  {Ziino}}, \bibinfo {author} {\bibfnamefont {W.}~\bibnamefont {Chang}},
  \bibinfo {author} {\bibfnamefont {S.~M.}\ \bibnamefont {Albrecht}}, \bibinfo
  {author} {\bibfnamefont {M.~H.}\ \bibnamefont {Madsen}}, \bibinfo {author}
  {\bibfnamefont {E.}~\bibnamefont {Johnson}}, \bibinfo {author} {\bibfnamefont
  {J.}~\bibnamefont {Nyg{\aa}rd}}, \bibinfo {author} {\bibfnamefont
  {C.}~\bibnamefont {Marcus}}, \ and\ \bibinfo {author} {\bibfnamefont {T.~S.}\
  \bibnamefont {Jespersen}},\ }\bibfield  {title} {\emph {\bibinfo {title}
  {Epitaxy of semiconductor-superconductor nanowires},\ }}\href {\doibase
  10.1038/nmat4176} {\bibfield  {journal} {\bibinfo  {journal} {Nat.~Mat.}\
  }\textbf {\bibinfo {volume} {14}},\ \bibinfo {pages} {400} (\bibinfo {year}
  {2015})}\BibitemShut {NoStop}%
\bibitem [{\citenamefont {{Chang}}\ \emph {et~al.}(2015)\citenamefont
  {{Chang}}, \citenamefont {{Albrecht}}, \citenamefont {{Jespersen}},
  \citenamefont {{Kuemmeth}}, \citenamefont {{Krogstrup}}, \citenamefont
  {{Nyg{\aa}rd}},\ and\ \citenamefont {{Marcus}}}]{Chang2015}%
  \BibitemOpen
  \bibfield  {author} {\bibinfo {author} {\bibfnamefont {W.}~\bibnamefont
  {{Chang}}}, \bibinfo {author} {\bibfnamefont {S.~M.}\ \bibnamefont
  {{Albrecht}}}, \bibinfo {author} {\bibfnamefont {T.~S.}\ \bibnamefont
  {{Jespersen}}}, \bibinfo {author} {\bibfnamefont {F.}~\bibnamefont
  {{Kuemmeth}}}, \bibinfo {author} {\bibfnamefont {P.}~\bibnamefont
  {{Krogstrup}}}, \bibinfo {author} {\bibfnamefont {J.}~\bibnamefont
  {{Nyg{\aa}rd}}}, \ and\ \bibinfo {author} {\bibfnamefont {C.~M.}\
  \bibnamefont {{Marcus}}},\ }\bibfield  {title} {\emph {\bibinfo {title} {Hard
  gap in epitaxial semiconductor-superconductor nanowires},\ }}\href {\doibase
  10.1038/nnano.2014.306} {\bibfield  {journal} {\bibinfo  {journal} {Nature
  Nanotechnology}\ }\textbf {\bibinfo {volume} {10}},\ \bibinfo {pages} {232}
  (\bibinfo {year} {2015})}\BibitemShut {NoStop}%
\bibitem [{\citenamefont {Deng}\ \emph {et~al.}(2016)\citenamefont {Deng},
  \citenamefont {Vaitiek}, \citenamefont {Hansen}, \citenamefont {Danon},
  \citenamefont {Leijnse}, \citenamefont {Flensberg}, \citenamefont
  {Krogstrup},\ and\ \citenamefont {Marcus}}]{Deng2016}%
  \BibitemOpen
  \bibfield  {author} {\bibinfo {author} {\bibfnamefont {M.~T.}\ \bibnamefont
  {Deng}}, \bibinfo {author} {\bibfnamefont {S.}~\bibnamefont {Vaitiek}},
  \bibinfo {author} {\bibfnamefont {E.~B.}\ \bibnamefont {Hansen}}, \bibinfo
  {author} {\bibfnamefont {J.}~\bibnamefont {Danon}}, \bibinfo {author}
  {\bibfnamefont {M.}~\bibnamefont {Leijnse}}, \bibinfo {author} {\bibfnamefont
  {K.}~\bibnamefont {Flensberg}}, \bibinfo {author} {\bibfnamefont
  {P.}~\bibnamefont {Krogstrup}}, \ and\ \bibinfo {author} {\bibfnamefont
  {C.~M.}\ \bibnamefont {Marcus}},\ }\bibfield  {title} {\emph {\bibinfo
  {title} {Majorana bound state in a coupled quantum-dot hybrid-nanowire
  system},\ }}\href {\doibase 10.1126/science.aaf3961} {\bibfield  {journal}
  {\bibinfo  {journal} {Science}\ }\textbf {\bibinfo {volume} {354}},\ \bibinfo
  {pages} {1557} (\bibinfo {year} {2016})}\BibitemShut {NoStop}%
\bibitem [{\citenamefont {Albrecht}\ \emph {et~al.}(2016)\citenamefont
  {Albrecht}, \citenamefont {Higginbotham}, \citenamefont {Madsen},
  \citenamefont {Kuemmeth}, \citenamefont {Jespersen}, \citenamefont
  {Nyg{\aa}rd}, \citenamefont {Krogstrup},\ and\ \citenamefont
  {Marcus}}]{Albrecht2016}%
  \BibitemOpen
  \bibfield  {author} {\bibinfo {author} {\bibfnamefont {S.~M.}\ \bibnamefont
  {Albrecht}}, \bibinfo {author} {\bibfnamefont {A.~P.}\ \bibnamefont
  {Higginbotham}}, \bibinfo {author} {\bibfnamefont {M.}~\bibnamefont
  {Madsen}}, \bibinfo {author} {\bibfnamefont {F.}~\bibnamefont {Kuemmeth}},
  \bibinfo {author} {\bibfnamefont {T.~S.}\ \bibnamefont {Jespersen}}, \bibinfo
  {author} {\bibfnamefont {J.}~\bibnamefont {Nyg{\aa}rd}}, \bibinfo {author}
  {\bibfnamefont {P.}~\bibnamefont {Krogstrup}}, \ and\ \bibinfo {author}
  {\bibfnamefont {C.~M.}\ \bibnamefont {Marcus}},\ }\bibfield  {title} {\emph
  {\bibinfo {title} {{Exponential Protection of Zero Modes in Majorana
  Islands}},\ }}\href {\doibase 10.1038/nature17162} {\bibfield  {journal}
  {\bibinfo  {journal} {Nature}\ }\textbf {\bibinfo {volume} {531}},\ \bibinfo
  {pages} {206} (\bibinfo {year} {2016})}\BibitemShut {NoStop}%
\bibitem [{\citenamefont {Nichele}\ \emph {et~al.}(2017)\citenamefont
  {Nichele}, \citenamefont {Drachmann}, \citenamefont {Whiticar}, \citenamefont
  {O'Farrell}, \citenamefont {Suominen}, \citenamefont {Fornieri},
  \citenamefont {Wang}, \citenamefont {Gardner}, \citenamefont {Thomas},
  \citenamefont {Hatke}, \citenamefont {Krogstrup}, \citenamefont {Manfra},
  \citenamefont {Flensberg},\ and\ \citenamefont {Marcus}}]{Nichele2017}%
  \BibitemOpen
  \bibfield  {author} {\bibinfo {author} {\bibfnamefont {F.}~\bibnamefont
  {Nichele}}, \bibinfo {author} {\bibfnamefont {A.~C.~C.}\ \bibnamefont
  {Drachmann}}, \bibinfo {author} {\bibfnamefont {A.~M.}\ \bibnamefont
  {Whiticar}}, \bibinfo {author} {\bibfnamefont {E.~C.~T.}\ \bibnamefont
  {O'Farrell}}, \bibinfo {author} {\bibfnamefont {H.~J.}\ \bibnamefont
  {Suominen}}, \bibinfo {author} {\bibfnamefont {A.}~\bibnamefont {Fornieri}},
  \bibinfo {author} {\bibfnamefont {T.}~\bibnamefont {Wang}}, \bibinfo {author}
  {\bibfnamefont {G.~C.}\ \bibnamefont {Gardner}}, \bibinfo {author}
  {\bibfnamefont {C.}~\bibnamefont {Thomas}}, \bibinfo {author} {\bibfnamefont
  {A.~T.}\ \bibnamefont {Hatke}}, \bibinfo {author} {\bibfnamefont
  {P.}~\bibnamefont {Krogstrup}}, \bibinfo {author} {\bibfnamefont {M.~J.}\
  \bibnamefont {Manfra}}, \bibinfo {author} {\bibfnamefont {K.}~\bibnamefont
  {Flensberg}}, \ and\ \bibinfo {author} {\bibfnamefont {C.~M.}\ \bibnamefont
  {Marcus}},\ }\bibfield  {title} {\emph {\bibinfo {title} {{Scaling of
  Majorana Zero-Bias Conductance Peaks}},\ }}\href {\doibase
  10.1103/PhysRevLett.119.136803} {\bibfield  {journal} {\bibinfo  {journal}
  {Phys.~Rev.~Lett.}\ }\textbf {\bibinfo {volume} {119}},\ \bibinfo {pages}
  {136803} (\bibinfo {year} {2017})}\BibitemShut {NoStop}%
\bibitem [{\citenamefont {Zhang}\ \emph {et~al.}(2018)\citenamefont {Zhang},
  \citenamefont {Liu}, \citenamefont {Gazibegovic}, \citenamefont {Xu},
  \citenamefont {Logan}, \citenamefont {Wang}, \citenamefont {van Loo},
  \citenamefont {Bommer}, \citenamefont {de~Moor}, \citenamefont {Car},
  \citenamefont {het Veld}, \citenamefont {van Veldhoven}, \citenamefont
  {Koelling}, \citenamefont {Verheijen}, \citenamefont {Pendharkar},
  \citenamefont {Pennachio}, \citenamefont {Shojaei}, \citenamefont {Lee},
  \citenamefont {Palmstrom}, \citenamefont {Bakkers}, \citenamefont {Sarma},\
  and\ \citenamefont {Kouwenhoven}}]{Zhang2018}%
  \BibitemOpen
  \bibfield  {author} {\bibinfo {author} {\bibfnamefont {H.}~\bibnamefont
  {Zhang}}, \bibinfo {author} {\bibfnamefont {C.-X.}\ \bibnamefont {Liu}},
  \bibinfo {author} {\bibfnamefont {S.}~\bibnamefont {Gazibegovic}}, \bibinfo
  {author} {\bibfnamefont {D.}~\bibnamefont {Xu}}, \bibinfo {author}
  {\bibfnamefont {J.~A.}\ \bibnamefont {Logan}}, \bibinfo {author}
  {\bibfnamefont {G.}~\bibnamefont {Wang}}, \bibinfo {author} {\bibfnamefont
  {N.}~\bibnamefont {van Loo}}, \bibinfo {author} {\bibfnamefont {J.~D.~S.}\
  \bibnamefont {Bommer}}, \bibinfo {author} {\bibfnamefont {M.~W.~A.}\
  \bibnamefont {de~Moor}}, \bibinfo {author} {\bibfnamefont {D.}~\bibnamefont
  {Car}}, \bibinfo {author} {\bibfnamefont {R.~L. M.~O.}\ \bibnamefont {het
  Veld}}, \bibinfo {author} {\bibfnamefont {P.~J.}\ \bibnamefont {van
  Veldhoven}}, \bibinfo {author} {\bibfnamefont {S.}~\bibnamefont {Koelling}},
  \bibinfo {author} {\bibfnamefont {M.~A.}\ \bibnamefont {Verheijen}}, \bibinfo
  {author} {\bibfnamefont {M.}~\bibnamefont {Pendharkar}}, \bibinfo {author}
  {\bibfnamefont {D.~J.}\ \bibnamefont {Pennachio}}, \bibinfo {author}
  {\bibfnamefont {B.}~\bibnamefont {Shojaei}}, \bibinfo {author} {\bibfnamefont
  {J.~S.}\ \bibnamefont {Lee}}, \bibinfo {author} {\bibfnamefont {C.~J.}\
  \bibnamefont {Palmstrom}}, \bibinfo {author} {\bibfnamefont {E.~P. A.~M.}\
  \bibnamefont {Bakkers}}, \bibinfo {author} {\bibfnamefont {S.~D.}\
  \bibnamefont {Sarma}}, \ and\ \bibinfo {author} {\bibfnamefont {L.~P.}\
  \bibnamefont {Kouwenhoven}},\ }\bibfield  {title} {\emph {\bibinfo {title}
  {{Quantized Majorana conductance}},\ }}\href {\doibase 10.1038/nature26142}
  {\bibfield  {journal} {\bibinfo  {journal} {Nature}\ }\textbf {\bibinfo
  {volume} {556}},\ \bibinfo {pages} {74} (\bibinfo {year} {2018})}\BibitemShut
  {NoStop}%
\bibitem [{\citenamefont {Suominen}\ \emph {et~al.}(2017)\citenamefont
  {Suominen}, \citenamefont {Kjaergaard}, \citenamefont {Hamilton},
  \citenamefont {Shabani}, \citenamefont {Palmstr\o{}m}, \citenamefont
  {Marcus},\ and\ \citenamefont {Nichele}}]{Suominen2017}%
  \BibitemOpen
  \bibfield  {author} {\bibinfo {author} {\bibfnamefont {H.~J.}\ \bibnamefont
  {Suominen}}, \bibinfo {author} {\bibfnamefont {M.}~\bibnamefont
  {Kjaergaard}}, \bibinfo {author} {\bibfnamefont {A.~R.}\ \bibnamefont
  {Hamilton}}, \bibinfo {author} {\bibfnamefont {J.}~\bibnamefont {Shabani}},
  \bibinfo {author} {\bibfnamefont {C.~J.}\ \bibnamefont {Palmstr\o{}m}},
  \bibinfo {author} {\bibfnamefont {C.~M.}\ \bibnamefont {Marcus}}, \ and\
  \bibinfo {author} {\bibfnamefont {F.}~\bibnamefont {Nichele}},\ }\bibfield
  {title} {\emph {\bibinfo {title} {{Zero-Energy Modes from Coalescing Andreev
  States in a Two-Dimensional Semiconductor-Superconductor Hybrid Platform}},\
  }}\href {\doibase 10.1103/PhysRevLett.119.176805} {\bibfield  {journal}
  {\bibinfo  {journal} {Phys.~Rev.~Lett.}\ }\textbf {\bibinfo {volume} {119}},\
  \bibinfo {pages} {176805} (\bibinfo {year} {2017})}\BibitemShut {NoStop}%
\bibitem [{\citenamefont {Vaitiekenas}\ \emph {et~al.}(2018)\citenamefont
  {Vaitiekenas}, \citenamefont {Deng}, \citenamefont {Nyg{\aa}rd},
  \citenamefont {Krogstrup},\ and\ \citenamefont {Marcus}}]{Vaitiekenas2018}%
  \BibitemOpen
  \bibfield  {author} {\bibinfo {author} {\bibfnamefont {S.}~\bibnamefont
  {Vaitiekenas}}, \bibinfo {author} {\bibfnamefont {M.~T.}\ \bibnamefont
  {Deng}}, \bibinfo {author} {\bibfnamefont {J.}~\bibnamefont {Nyg{\aa}rd}},
  \bibinfo {author} {\bibfnamefont {P.}~\bibnamefont {Krogstrup}}, \ and\
  \bibinfo {author} {\bibfnamefont {C.~M.}\ \bibnamefont {Marcus}},\ }\bibfield
   {title} {\emph {\bibinfo {title} {{Effective g Factor of Subgap States in
  Hybrid Nanowires}},\ }}\href {\doibase 10.1103/PhysRevLett.121.037703}
  {\bibfield  {journal} {\bibinfo  {journal} {Phys.~Rev.~Lett.}\ }\textbf
  {\bibinfo {volume} {121}},\ \bibinfo {pages} {037703} (\bibinfo {year}
  {2018})}\BibitemShut {NoStop}%
\bibitem [{\citenamefont {Winkler}\ \emph {et~al.}(2017)\citenamefont
  {Winkler}, \citenamefont {Varjas}, \citenamefont {Skolasinski}, \citenamefont
  {Soluyanov}, \citenamefont {Troyer},\ and\ \citenamefont
  {Wimmer}}]{Winkler2017}%
  \BibitemOpen
  \bibfield  {author} {\bibinfo {author} {\bibfnamefont {G.~W.}\ \bibnamefont
  {Winkler}}, \bibinfo {author} {\bibfnamefont {D.}~\bibnamefont {Varjas}},
  \bibinfo {author} {\bibfnamefont {R.}~\bibnamefont {Skolasinski}}, \bibinfo
  {author} {\bibfnamefont {A.~A.}\ \bibnamefont {Soluyanov}}, \bibinfo {author}
  {\bibfnamefont {M.}~\bibnamefont {Troyer}}, \ and\ \bibinfo {author}
  {\bibfnamefont {M.}~\bibnamefont {Wimmer}},\ }\bibfield  {title} {\emph
  {\bibinfo {title} {{Orbital Contributions to the Electron $g$ Factor in
  Semiconductor Nanowires}},\ }}\href {\doibase 10.1103/PhysRevLett.119.037701}
  {\bibfield  {journal} {\bibinfo  {journal} {Phys.~Rev.~Lett.}\ }\textbf
  {\bibinfo {volume} {119}},\ \bibinfo {pages} {037701} (\bibinfo {year}
  {2017})}\BibitemShut {NoStop}%
\bibitem [{\citenamefont {Vuik}\ \emph {et~al.}(2016)\citenamefont {Vuik},
  \citenamefont {Eeltink}, \citenamefont {Akhmerov},\ and\ \citenamefont
  {Wimmer}}]{Vuik2016}%
  \BibitemOpen
  \bibfield  {author} {\bibinfo {author} {\bibfnamefont {A.}~\bibnamefont
  {Vuik}}, \bibinfo {author} {\bibfnamefont {D.}~\bibnamefont {Eeltink}},
  \bibinfo {author} {\bibfnamefont {A.~R.}\ \bibnamefont {Akhmerov}}, \ and\
  \bibinfo {author} {\bibfnamefont {M.}~\bibnamefont {Wimmer}},\ }\bibfield
  {title} {\emph {\bibinfo {title} {{Effects of the electrostatic environment
  on the Majorana nanowire devices}},\ }}\href {\doibase
  10.1088/1367-2630/18/3/033013} {\bibfield  {journal} {\bibinfo  {journal}
  {New Journal of Physics}\ }\textbf {\bibinfo {volume} {18}},\ \bibinfo
  {pages} {033013} (\bibinfo {year} {2016})}\BibitemShut {NoStop}%
\bibitem [{\citenamefont {Dominguez}\ \emph {et~al.}(2017)\citenamefont
  {Dominguez}, \citenamefont {Cayao}, \citenamefont {San-Jose}, \citenamefont
  {Aguado}, \citenamefont {Yeyati},\ and\ \citenamefont
  {Prada}}]{Dominguez2017}%
  \BibitemOpen
  \bibfield  {author} {\bibinfo {author} {\bibfnamefont {F.}~\bibnamefont
  {Dominguez}}, \bibinfo {author} {\bibfnamefont {J.}~\bibnamefont {Cayao}},
  \bibinfo {author} {\bibfnamefont {P.}~\bibnamefont {San-Jose}}, \bibinfo
  {author} {\bibfnamefont {R.}~\bibnamefont {Aguado}}, \bibinfo {author}
  {\bibfnamefont {A.~L.}\ \bibnamefont {Yeyati}}, \ and\ \bibinfo {author}
  {\bibfnamefont {E.}~\bibnamefont {Prada}},\ }\bibfield  {title} {\emph
  {\bibinfo {title} {{Zero-energy pinning from interactions in Majorana
  nanowires}},\ }}\href {\doibase 10.1038/s41535-017-0012-0} {\bibfield
  {journal} {\bibinfo  {journal} {npj Quantum Materials}\ }\textbf {\bibinfo
  {volume} {2}},\ \bibinfo {pages} {13} (\bibinfo {year} {2017})}\BibitemShut
  {NoStop}%
\bibitem [{Esc()}]{Escribano}%
  \BibitemOpen
  \href@noop {} {}\bibinfo {note} {S.D. Escribano, A. Levy Yeyati, and E.
  Prada, \href{https://arxiv.org/abs/1712.07625}{arXiv:1712.07625}}\BibitemShut
  {NoStop}%
\bibitem [{\citenamefont {{van Heck}}\ \emph {et~al.}(2016)\citenamefont {{van
  Heck}}, \citenamefont {Lutchyn},\ and\ \citenamefont
  {Glazman}}]{VanHeck2016}%
  \BibitemOpen
  \bibfield  {author} {\bibinfo {author} {\bibfnamefont {B.}~\bibnamefont {{van
  Heck}}}, \bibinfo {author} {\bibfnamefont {R.~M.}\ \bibnamefont {Lutchyn}}, \
  and\ \bibinfo {author} {\bibfnamefont {L.~I.}\ \bibnamefont {Glazman}},\
  }\bibfield  {title} {\emph {\bibinfo {title} {{Conductance of a proximitized
  nanowire in the Coulomb blockade regime}},\ }}\href {\doibase
  10.1103/PhysRevB.93.235431} {\bibfield  {journal} {\bibinfo  {journal}
  {Phys.~Rev.~B}\ }\textbf {\bibinfo {volume} {93}},\ \bibinfo {pages} {235431}
  (\bibinfo {year} {2016})}\BibitemShut {NoStop}%
\bibitem [{\citenamefont {Reeg}\ \emph {et~al.}(2017)\citenamefont {Reeg},
  \citenamefont {Loss},\ and\ \citenamefont {Klinovaja}}]{Reeg2017}%
  \BibitemOpen
  \bibfield  {author} {\bibinfo {author} {\bibfnamefont {C.}~\bibnamefont
  {Reeg}}, \bibinfo {author} {\bibfnamefont {D.}~\bibnamefont {Loss}}, \ and\
  \bibinfo {author} {\bibfnamefont {J.}~\bibnamefont {Klinovaja}},\ }\bibfield
  {title} {\emph {\bibinfo {title} {{Finite-size effects in a nanowire strongly
  coupled to a thin superconducting shell}},\ }}\href {\doibase
  10.1103/PhysRevB.96.125426} {\bibfield  {journal} {\bibinfo  {journal}
  {Phys.~Rev.~B}\ }\textbf {\bibinfo {volume} {96}},\ \bibinfo {pages} {125426}
  (\bibinfo {year} {2017})}\BibitemShut {NoStop}%
\bibitem [{\citenamefont {Stanescu}\ and\ \citenamefont {{Das
  Sarma}}(2017)}]{Stanescu2017}%
  \BibitemOpen
  \bibfield  {author} {\bibinfo {author} {\bibfnamefont {T.~D.}\ \bibnamefont
  {Stanescu}}\ and\ \bibinfo {author} {\bibfnamefont {S.}~\bibnamefont {{Das
  Sarma}}},\ }\bibfield  {title} {\emph {\bibinfo {title} {{Proximity-induced
  low-energy renormalization in hybrid semiconductor-superconductor Majorana
  structures}},\ }}\href {\doibase 10.1103/PhysRevB.96.014510} {\bibfield
  {journal} {\bibinfo  {journal} {Phys.~Rev.~B}\ }\textbf {\bibinfo {volume}
  {96}},\ \bibinfo {pages} {014510} (\bibinfo {year} {2017})}\BibitemShut
  {NoStop}%
\bibitem [{\citenamefont {Rainis}\ and\ \citenamefont
  {Loss}(2014)}]{Rainis2014}%
  \BibitemOpen
  \bibfield  {author} {\bibinfo {author} {\bibfnamefont {D.}~\bibnamefont
  {Rainis}}\ and\ \bibinfo {author} {\bibfnamefont {D.}~\bibnamefont {Loss}},\
  }\bibfield  {title} {\emph {\bibinfo {title} {{Conductance behavior in
  nanowires with spin-orbit interaction: A numerical study}},\ }}\href
  {\doibase 10.1103/PhysRevB.90.235415} {\bibfield  {journal} {\bibinfo
  {journal} {Phys.~Rev.~B}\ }\textbf {\bibinfo {volume} {90}},\ \bibinfo
  {pages} {235415} (\bibinfo {year} {2014})}\BibitemShut {NoStop}%
\bibitem [{\citenamefont {Prada}\ \emph {et~al.}(2017)\citenamefont {Prada},
  \citenamefont {Aguado},\ and\ \citenamefont {San-Jose}}]{Prada2017}%
  \BibitemOpen
  \bibfield  {author} {\bibinfo {author} {\bibfnamefont {E.}~\bibnamefont
  {Prada}}, \bibinfo {author} {\bibfnamefont {R.}~\bibnamefont {Aguado}}, \
  and\ \bibinfo {author} {\bibfnamefont {P.}~\bibnamefont {San-Jose}},\
  }\bibfield  {title} {\emph {\bibinfo {title} {{Measuring Majorana nonlocality
  and spin structure with a quantum dot}},\ }}\href {\doibase
  10.1103/PhysRevB.96.085418} {\bibfield  {journal} {\bibinfo  {journal}
  {Phys.~Rev.~B}\ }\textbf {\bibinfo {volume} {96}},\ \bibinfo {pages} {085418}
  (\bibinfo {year} {2017})}\BibitemShut {NoStop}%
\bibitem [{Sau()}]{Saulius}%
  \BibitemOpen
  \href@noop {} {}\bibinfo {note} {S. {Vaitiek{\.e}nas}, M.~T. {Deng}, J.
  {Nyg{\aa}rd}, P. {Krogstrup}, C.M. {Marcus}, \textit{Effective g-factor in
  Majorana Wires},
  \href{https://arxiv.org/abs/1710.04300}{arXiv:1710.04300}}\BibitemShut
  {NoStop}%
\bibitem [{Gat()}]{GateScaling}%
  \BibitemOpen
  \href@noop {} {}\bibinfo {note} {By employing the boundary conditions for the
  electric field ($\bm{{\cal E}}$) across the semiconductor-dielectric
  boundary, one finds that $V_\G=V_\D-L_\D{\cal
  E}_z(L_2^-)\varepsilon_\InAs/\varepsilon_\D$, with $L_\D$ denoting the
  thickness of the dielectric and $\varepsilon_\D$ its dielectric constant. For
  an illustration, we calculate $V_\G$ in the case of $V_\D=-0.5\ph \V$ and
  $V_\D=0.1\ph \V$ of Fig.~\ref{fig:DeviceSketch}(d), for a choice of a
  dielectric layer as in Ref.~\onlinecite{Saulius}. The electric field in the
  former (latter) case is ${\cal E}_z(L_2^-)\simeq 4.8\ph{\rm mV}\nm^{-1}$
  (${\cal E}_z(L_2^-)\simeq -12.3\ph{\rm mV}\nm^{-1}$). For a $L_\D=10\ph\nm$
  thick HfO$_2$ layer, in which case $\varepsilon_\D=25$, we obtain the
  correspondence $V_\D=\{-0.5\ph V,0.1\ph V\}\mapsto V_\G\approx\{-0.53\ph
  V,0.18\ph V\}$}\BibitemShut {NoStop}%
\bibitem [{\citenamefont {Ashcroft}\ and\ \citenamefont
  {Mermin}(1976)}]{Ashcroft}%
  \BibitemOpen
  \bibfield  {author} {\bibinfo {author} {\bibfnamefont {N.}~\bibnamefont
  {Ashcroft}}\ and\ \bibinfo {author} {\bibfnamefont {N.}~\bibnamefont
  {Mermin}},\ }\href@noop {} {\emph {\bibinfo {title} {{Solid State
  Physics}}}}\ (\bibinfo  {publisher} {Saunders College},\ \bibinfo {address}
  {Philadelphia},\ \bibinfo {year} {1976})\BibitemShut {NoStop}%
\bibitem [{\citenamefont {Winkler}(2003)}]{Winkler}%
  \BibitemOpen
  \bibfield  {author} {\bibinfo {author} {\bibfnamefont {R.}~\bibnamefont
  {Winkler}},\ }\href@noop {} {\emph {\bibinfo {title} {{Spin-orbit coupling
  effects in two-dimensional electron and hole systems}}}},\ Springer tracts in
  modern physics\ (\bibinfo  {publisher} {Springer},\ \bibinfo {address}
  {Berlin},\ \bibinfo {year} {2003})\BibitemShut {NoStop}%
\bibitem [{ARP()}]{ARPES}%
  \BibitemOpen
  \href@noop {} {}\bibinfo {note} {Recent ARPES measurements on planar
  epitaxial InAs/Al interfaces with an Zincblende InAs (100) have revealed an
  interfacial band offset between the InAs conductance band and the Fermi level
  of $\Phi=0.23$ eV, S.~Schuwalow, P.~Krogstrup et al., to be
  published.}\BibitemShut {Stop}%
\bibitem [{\citenamefont {Eastment}\ and\ \citenamefont
  {Mee}(1973)}]{AlWorkfunction}%
  \BibitemOpen
  \bibfield  {author} {\bibinfo {author} {\bibfnamefont {R.~M.}\ \bibnamefont
  {Eastment}}\ and\ \bibinfo {author} {\bibfnamefont {C.~H.~B.}\ \bibnamefont
  {Mee}},\ }\bibfield  {title} {\emph {\bibinfo {title} {Work function
  measurements on (100), (110) and (111) surfaces of aluminium},\ }}\href
  {http://stacks.iop.org/0305-4608/3/i=9/a=016} {\bibfield  {journal} {\bibinfo
   {journal} {Journal of Physics F: Metal Physics}\ }\textbf {\bibinfo {volume}
  {3}},\ \bibinfo {pages} {1738} (\bibinfo {year} {1973})}\BibitemShut
  {NoStop}%
\bibitem [{\citenamefont {Gobeli}\ and\ \citenamefont
  {Allen}(1965)}]{InAsElectronAffinity}%
  \BibitemOpen
  \bibfield  {author} {\bibinfo {author} {\bibfnamefont {G.~W.}\ \bibnamefont
  {Gobeli}}\ and\ \bibinfo {author} {\bibfnamefont {F.~G.}\ \bibnamefont
  {Allen}},\ }\bibfield  {title} {\emph {\bibinfo {title} {{Photoelectric
  Properties of Cleaved GaAs, GaSb, InAs, and InSb Surfaces; Comparison with Si
  and Ge}},\ }}\href {\doibase 10.1103/PhysRev.137.A245} {\bibfield  {journal}
  {\bibinfo  {journal} {Phys. Rev.}\ }\textbf {\bibinfo {volume} {137}},\
  \bibinfo {pages} {A245} (\bibinfo {year} {1965})}\BibitemShut {NoStop}%
\bibitem [{\citenamefont {Cheng}\ \emph {et~al.}(2009)\citenamefont {Cheng},
  \citenamefont {Lutchyn}, \citenamefont {Galitski},\ and\ \citenamefont
  {DasSarma}}]{Cheng2009}%
  \BibitemOpen
  \bibfield  {author} {\bibinfo {author} {\bibfnamefont {M.}~\bibnamefont
  {Cheng}}, \bibinfo {author} {\bibfnamefont {R.~M.}\ \bibnamefont {Lutchyn}},
  \bibinfo {author} {\bibfnamefont {V.}~\bibnamefont {Galitski}}, \ and\
  \bibinfo {author} {\bibfnamefont {S.}~\bibnamefont {DasSarma}},\ }\bibfield
  {title} {\emph {\bibinfo {title} {{Splitting of Majorana-Fermion Modes due to
  Intervortex Tunneling in a px+ipy Superconductor}},\ }}\href {\doibase
  10.1103/PhysRevLett.103.107001} {\bibfield  {journal} {\bibinfo  {journal}
  {Phys.~Rev.~Lett.}\ }\textbf {\bibinfo {volume} {103}},\ \bibinfo {pages}
  {107001} (\bibinfo {year} {2009})}\BibitemShut {NoStop}%
\bibitem [{\citenamefont {Rainis}\ \emph {et~al.}(2013)\citenamefont {Rainis},
  \citenamefont {Trifunovic}, \citenamefont {Klinovaja},\ and\ \citenamefont
  {Loss}}]{Rainis2013}%
  \BibitemOpen
  \bibfield  {author} {\bibinfo {author} {\bibfnamefont {D.}~\bibnamefont
  {Rainis}}, \bibinfo {author} {\bibfnamefont {L.}~\bibnamefont {Trifunovic}},
  \bibinfo {author} {\bibfnamefont {J.}~\bibnamefont {Klinovaja}}, \ and\
  \bibinfo {author} {\bibfnamefont {D.}~\bibnamefont {Loss}},\ }\bibfield
  {title} {\emph {\bibinfo {title} {{Towards a realistic transport modeling in
  a superconducting nanowire with Majorana fermions}},\ }}\href {\doibase
  10.1103/PhysRevB.87.024515} {\bibfield  {journal} {\bibinfo  {journal}
  {Phys.~Rev.~B}\ }\textbf {\bibinfo {volume} {87}},\ \bibinfo {pages} {024515}
  (\bibinfo {year} {2013})}\BibitemShut {NoStop}%
\bibitem [{\citenamefont {DasSarma}\ \emph {et~al.}(2012)\citenamefont
  {DasSarma}, \citenamefont {Sau},\ and\ \citenamefont
  {Stanescu}}]{DasSarma2012}%
  \BibitemOpen
  \bibfield  {author} {\bibinfo {author} {\bibfnamefont {S.}~\bibnamefont
  {DasSarma}}, \bibinfo {author} {\bibfnamefont {J.~D.}\ \bibnamefont {Sau}}, \
  and\ \bibinfo {author} {\bibfnamefont {T.~D.}\ \bibnamefont {Stanescu}},\
  }\bibfield  {title} {\emph {\bibinfo {title} {{Splitting of the zero-bias
  conductance peak as smoking gun evidence for the existence of the Majorana
  mode in a superconductor-semiconductor nanowire}},\ }}\href {\doibase
  10.1103/PhysRevB.86.220506} {\bibfield  {journal} {\bibinfo  {journal}
  {Phys.~Rev.~B}\ }\textbf {\bibinfo {volume} {86}},\ \bibinfo {pages} {220506}
  (\bibinfo {year} {2012})}\BibitemShut {NoStop}%
\bibitem [{\citenamefont {Mishmash}\ \emph {et~al.}(2016)\citenamefont
  {Mishmash}, \citenamefont {Aasen}, \citenamefont {Higginbotham},\ and\
  \citenamefont {Alicea}}]{Mishmash2016}%
  \BibitemOpen
  \bibfield  {author} {\bibinfo {author} {\bibfnamefont {R.~V.}\ \bibnamefont
  {Mishmash}}, \bibinfo {author} {\bibfnamefont {D.}~\bibnamefont {Aasen}},
  \bibinfo {author} {\bibfnamefont {A.~P.}\ \bibnamefont {Higginbotham}}, \
  and\ \bibinfo {author} {\bibfnamefont {J.}~\bibnamefont {Alicea}},\
  }\bibfield  {title} {\emph {\bibinfo {title} {{Approaching a topological
  phase transition in Majorana nanowires}},\ }}\href {\doibase
  10.1103/PhysRevB.93.245404} {\bibfield  {journal} {\bibinfo  {journal}
  {Phys.~Rev.~B}\ }\textbf {\bibinfo {volume} {93}},\ \bibinfo {pages} {245404}
  (\bibinfo {year} {2016})}\BibitemShut {NoStop}%
\bibitem [{Ant()}]{Antipov2018}%
  \BibitemOpen
  \href@noop {} {}\bibinfo {note} {A.~E. Antipov \textit{et. al.},
  \textit{Effects of gate-induced electric fields on semiconductor Majorana
  nanowires},
  \href{https://arxiv.org/abs/1801.02616}{arXiv:1801.02616}.}\BibitemShut
  {Stop}%
\bibitem [{Woo()}]{Woods2018}%
  \BibitemOpen
  \href@noop {} {}\bibinfo {note} {B.~D. Woods A. \textit{et. al.},
  \textit{Effective theory approach to the Schrodinger-Poisson problem in
  semiconductor Majorana devices},
  \href{https://arxiv.org/abs/1801.02630}{arXiv:1801.02630}}\BibitemShut
  {NoStop}%
\end{thebibliography}
\end{document}